\documentclass[11pt]{article}
\usepackage{geometry,latexsym,amssymb,float}
\usepackage{tikz}
\usetikzlibrary{shapes,arrows}
\usepackage{enumerate}
\newtheorem{theorem}{Theorem}

\newtheorem{observation}[theorem]{Observation}
\newtheorem{lemma}[theorem]{Lemma}
\newtheorem{corollary}[theorem]{Corollary}

\newtheorem{claim}{Claim}

\def \proof {\noindent{\bf Proof}: }

\usepackage{setspace}
\begin{document}
\parskip .3cm
\parindent 0cm
\large

\begin{center}
\Large
{\bf Perfect Edge Domination: Hard and Solvable Cases} \\
\vspace{1.5cm}

\large Min Chih Lin$^1$ \hspace{2cm} Vadim Lozin$^2$ \\ Veronica
A. Moyano$^1$ \hspace{2cm} Jayme L Szwarcfiter$^3$
\bigskip

\normalsize
$^1$ Universidad de Buenos Aires \\
Argentina \\
oscarlin@dc.uba.ar, vmoyano@ic.fcen.uba.ar \\
\bigskip

$^2$ University of Warwick \\
United Kingdom \\
v.lozin@warwick.ac.uk \\
\bigskip

$^3$ Universidade Federal do Rio de Janeiro, and \\
Instituto Nacional de Metrologia, Qualidade e Tecnologia \\
Brasil \\
jayme@nce.ufrj.br \\
\vspace{2cm}

ABSTRACT \\
\end{center}

\small Let $G$ be an undirected graph. An edge of $G$ {\it dominates}
itself and all edges adjacent to it. A subset $E'$ of edges of $G$
is an {\it edge dominating set} of $G$, if every edge of the
graph is dominated by some edge of $E'$. We say that $E'$ is a {\it
perfect edge dominating set} of $G$, if every edge not in $E'$
is dominated by exactly one edge of $E'$. The {\it perfect edge
dominating problem} is to determine a least cardinality perfect edge
dominating set of $G$. For this problem,  we describe two
NP-completeness proofs, for the classes of claw-free graphs of
degree at most 3, and for bounded degree graphs, of maximum
degree at most $d \geq 3$ and large girth. In
contrast, we prove that the problem admits an $O(n)$ time
solution, for cubic claw-free graphs. 
In addition, we prove a complexity dichotomy theorem for the
perfect edge domination problem, based on the results described in the
paper. Finally, we describe a linear time algorithm for finding a
minimum weight perfect edge dominating set of a $P_5$-free graph. The
algorithm is robust, in the sense that, given an arbitrary graph $G$,
either it computes a minimum weight perfect edge dominating set of
$G$, or it exhibits an induced subgraph of $G$, isomorphic to a $P_5$.

\section{Introduction}
Edge domination problems have been the focus of
considerable attention, in the last few years. Among the relevant variations of these problems are the perfect edge domination and
efficient edge domination. In the former variation, each edge not
in the dominating set is dominated by exactly one edge, while in
the latter, every edge of the graph is so dominated. The problems
consist in determining such dominating sets, of minimum
cardinality or minimum total weight of their edges. Most of the
known results, so far concerned the efficient edge dominating case (also known
as {\it dominating induced matching}), for instance see \cite{Br-Le-Ra,Br-Mo,Ca-Ce-De-Si,Ca-Ko-Lo,Gr-Sl-Sh-Ho,He-Lo-Ri-Za-We,Li-Mi-Sz-13,Li-Mi-Sz-13a,Li-Mi-Sz-14,Li-Mi-Sz-15,Li-Mo-Ra-Sz,Lu-Ko-Ta,Lu-Ta,Xi-Na}. These two types of
dominations may lead to problems of a quite different nature. To
start with, an efficient edge dominating set of a graph may not
exist, while it necessarily exists for the perfect domination.
Furthermore, there are important differences concerning the
complexity status of the problems for classes of graphs although
both of them are hard, in general. In fact, perfect edge
domination problems seem to be not easier to tackle than efficient edge
domination. On the other hand, if a graph contains an efficient
edge dominating set then such set is also a minimum cardinality
perfect edge dominating set \cite{Ge-Ha-Sa-Wh},\cite{Lu-Ko-Ta}.
The corresponding decision problems, for the cardinality version, are therefore as follows.\\

EFFICIENT EDGE DOMINATION \\
INPUT: Graph $G$\\
QUESTION: Does $G$ contain an efficient edge dominating set ?\\

PERFECT EDGE DOMINATION \\
INPUT: Graph $G$, integer $p$ \\
QUESTION: Does $G$ contain a perfect edge dominating set S of size
$\leq p$ ?\\\\
Both problems are known to be NP-complete \cite{Gr-Sl-Sh-Ho} and
\cite{Lu-Ko-Ta}, respectively.

In the present paper, we consider perfect edge domination.
The main proposed results are as follows:

\begin{itemize}
\item  A NP-hardness proof for the cardinality version of
perfect edge domination in claw-free graphs of degree at most 3.
\item A linear time solution for weighted (with possibly negative weights) perfect edge domination in cubic claw-free graphs. 
\item A NP-hardness proof for the
cardinality version of perfect edge domination in bounded-degree graphs of large girth. The proof also implies NP-hardness for
$r$-regular graphs, $r \geq 3$. 
\item A complexity dichotomy theorem which establishes NP-hardness for the perfect edge domination problem in any class of bounded degree graphs, defined by forbidding one general graph $H$. The only exception is when $H$ is a set of disjoint paths, in which case the corresponding perfect edge domination problem admits a polynomial time solution.
\item A robust linear time algorithm for solving weighted perfect
edge domination problem of $P_5$-free graphs. That is, in linear
time, the algorithm either finds a minimum weight edge
dominating set of the graph, or exhibits an induced $P_5$. 
\end{itemize}

We remark that the above problems have already been solved for
efficient edge domination. In contrast with the perfect edge
domination hardness proposed in the present paper, Cardoso,
Koperlainen and Lozin \cite{Ca-Ko-Lo} have described an $O(n^2)$
time algorithm for finding an efficient edge dominating set, for general claw-free graphs. More
recently, the corresponding weighted problem has been solved in
$O(n)$ time, by Lin, Mizrahi and Szwarcfiter \cite{Li-Mi-Sz-14}.
Moreover, Hertz, Lozin, Ries, Zamaraev and de Werra
\cite{He-Lo-Ri-Za-We} have shown that the efficient edge
domination problem can still be solved in polynomial time for
graphs containing no long claws. Finally, for bounded degree
graphs, Cardoso, Cerdeira, Delorme and Silva \cite{Ca-Ce-De-Si}
have shown that efficient edge domination problem is NP-complete for
$r$-regular graphs for $r\geq3$.

Known results to the authors for the perfect edge domination
problem are as follows. Lu, Ko and Tang \cite{Lu-Ko-Ta} proved the
problem is NP-complete for bipartite graphs. As for polynomial
time solvable cases, there are linear time algorithms also
described by Lu, Ko and Tang,  for generalized series-parallel
graphs and chordal graphs. In addition, there is a linear time algorithm
for circular-arc graphs, by Lin, Mizrahi and Szwarcfiter
\cite{Li-Mi-Sz-15}. All of these algorithms solve the weighted
perfect edge domination problem.

The structure of the paper is as follow. Section 2 describes 
the terminology employed. Section 3 contains the NP-completeness proof 
of the perfect edge domination problem for claw-free graphs
of degree at most three. In Section 4 we show that the weighted perfect edge 
domination problem is solvable in linear time for cubic claw-free graphs.
In Section 5 we prove NP-hardness of the perfect edge domination 
problem in bounded degree graphs even restricted to those of large girth.
This result implies the NP-hardness for $r$-regular graphs with $r\geq 3$.
Section 6 contains the dichotomy theorem for the complexity of the perfect edge 
domination problem for bounded degree $H$-free graphs. When $H$ is a disjoint union of paths
the problem is solvable in polynomial time, otherwise it remains NP-hard.
Section 7 is dedicated to the case $H=P_k$. For $P_5$-free graphs we present a robust algorithm
which solves the weighted perfect edge domination problem in linear time.

\section{Preliminaries}

Let $G$ be an undirected graph with no loops or multiple
edges. The vertex and edge sets of $G$ are denoted by $V(G)$ and
$E(G)$, respectively, $|V(G)| = n$. For $v,w$, adjacent vertices of
$G$, write $vw$ to denote the edge incident to $v$ and $w$. For $v
\in V(G)$, let $N(v) = \{w \in V(G) | v,w$ are adjacent$\}$, and
$N[v] = N(v) \cup \{v\}$. The edges $e$ and $e'$ are adjacent if they
are incident to a common vertex. For $e \in E(G)$, let $N'(e) = \{e' \in
E(G) | e,e'$ are adjacent$\}$, and $N'[e] = N'(e) \cup \{e\}$. For
$V' \subseteq V(G)$, denote by $G[V']$ the subgraph induced in $G$
by $V'$. Similarly, for $E' \subseteq E(G)$, denote by $G[E']$ the
subgraph of $G$ having exactly the edges of $E'$ without isolated vertices. 
Each edge $e \in E(G)$ may be assigned a real value, called the
{\it weight} of $e$.
The girth of $G$ is the length of a shortest cycle contained in $G$.
If $G$ is an acycle graph the girth is defined to be infinity.

A vertex $v \in V(G)$ {\it dominates} itself and any other vertex
adjacent to it. A subset of vertices $X \subseteq V(G)$ is a {\it
(vertex) dominating set} if every vertex of $G$ is dominated by
some vertex of $X$.  An edge $vw \in E(G)$ {\it dominates} itself
and any other edge adjacent to it. A subset of edges $E' \subseteq
E(G)$ is a {\it perfect edge dominating set} (PEDS) of $G$, if every edge
of $E(G) \setminus E'$ is dominated by exactly one edge of $E'$.
On the other hand if every edge of $E(G)$ is dominated exactly once by $E'$
then $E'$ is an {\it efficient edge dominating set} (EEDS), also called a
{\it dominating induced matching}. The {\it cardinality perfect
(efficient) edge domination problem} is to determine the perfect
(efficient) edge dominating set of $G$ of least cardinality. The
corresponding {\it weighted} problems are defined replacing
minimum cardinality by minimum sum of weights of the dominating
edges.

Let $P \subseteq E(G)$ be a perfect edge dominating set of a
connected graph $G$. Then $P$ defines a 3-coloring  of the
vertices of $G$,  as below:

\begin{itemize}
\item {\it black vertices}: Those having at least two incident
edges of $P$. We denote this subset of vertices by $B$.
\item {\it yellow vertices}: Those which are incident to exactly
one edge of $P$. We denote this subset of vertices by $Y$.
\item {\it white vertices}: The ones not incident to any edge
of $P$. We denote this subset of vertices by $W$.
\end{itemize}

We call $(B,Y,W)$ the \emph{3-coloring associated to $P$}.

\begin{observation}\label{O1}
$W$ is an independent set.
\end{observation}

\begin{observation}\label{O2}
Pendant vertices (vertices of degree 1) in $G\setminus W$ are
exactly the yellow vertices.
\end{observation}

\begin{observation}\label{O3}
White vertices have only yellow neighbors.
\end{observation}

\begin{observation}\label{O4}
Every vertex of an induced $K_t$ with $t\geq 4$ must be black.
\end{observation}

\begin{observation}\label{O5}
Every induced triangle has (i) three black vertices or (ii) two
yellow vertices and one white vertex.
\end{observation}

It is straightforward to see that any 3-coloring $(B,Y,W)$ of the
vertices of $G$, that verifies the conditions of Observations
\ref{O1}, \ref{O2} and \ref{O3}, is
necessarily associated to some perfect edge dominating set of $G$.\\

On the other hand, an efficient edge dominating set is clearly a
perfect edge dominating set. In this case, the set $B$ is empty.
The latter implies that the subset induced by $Y$ is an induced
matching of $G$.

Clearly,  the converse is not true. When $B \not=
\emptyset$,  $P$ is not an efficient edge dominating set. In
particular, $E(G)$ is a perfect edge dominating set, called the
{\it trivial perfect edge dominating set}, when $W=\emptyset$. Any perfect edge dominating set
$P$ which is neither trivial nor an efficient edge dominating set
is called {\it proper perfect edge dominating set} because $W, B,
Y\neq \emptyset$. Similarly,
$(B,Y,W)$ is then called a {\it proper coloring}.

\section{Claw-free graphs of degree at most three}

In this section, we prove NP-completeness of the perfect edge
domination problem, for claw-free graphs, of degree at most three.
We recall that the efficient edge domination problem can be
solved in polynomial time for general claw-free graphs \cite{Ca-Ko-Lo,Li-Mi-Sz-14}. However, for $r$-regular graphs, fixed $r \geq 3$,
efficient edge domination problem is also NP-complete \cite{Ca-Ce-De-Si}.

\begin{theorem}\cite{Ca-Ce-De-Si}\label{T1}
For arbitrary fixed $r\geq 3$, deciding on the existence of
efficient edge dominating sets on $r$-regular graphs is
NP-complete.
\end{theorem}

\begin{figure}
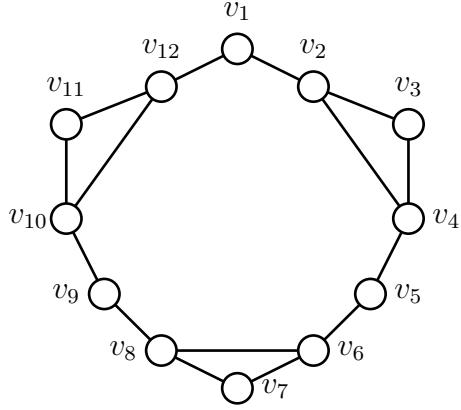

\begin{center}
\centering
\begin{pgfpicture}{0cm}{0cm}{12cm}{6cm}
    \pgfsetlinewidth{1pt}

    \pgfnodecircle{v1}[stroke]{\pgfxy(0+6,4.5+1.5-1)}{2mm}
    \pgfnodecircle{v2}[stroke]{\pgfxy(1+6,4+1.5-1)}{2mm}
    \pgfnodecircle{v3}[stroke]{\pgfxy(2.25+6,3.5+1.5-1)}{2mm}
    \pgfnodecircle{v4}[stroke]{\pgfxy(2.25+6,2.25+1.5-1)}{2mm}
    \pgfnodecircle{v5}[stroke]{\pgfxy(1.75+6,1.25+1.5-1)}{2mm}
    \pgfnodecircle{v6}[stroke]{\pgfxy(1+6,0.5+1.5-1)}{2mm}
    \pgfnodecircle{v7}[stroke]{\pgfxy(0+6,0+1.5-1)}{2mm}
    \pgfnodecircle{v8}[stroke]{\pgfxy(-1+6,0.5+1.5-1)}{2mm}
    \pgfnodecircle{v9}[stroke]{\pgfxy(-1.75+6,1.25+1.5-1)}{2mm}
    \pgfnodecircle{v10}[stroke]{\pgfxy(-2.25+6,2.25+1.5-1)}{2mm}
    \pgfnodecircle{v11}[stroke]{\pgfxy(-2.25+6,3.5+1.5-1)}{2mm}
    \pgfnodecircle{v12}[stroke]{\pgfxy(-1+6,4+1.5-1)}{2mm}
    \pgfnodeconnline{v1}{v2}
    \pgfnodeconnline{v2}{v3}
    \pgfnodeconnline{v3}{v4}
    \pgfnodeconnline{v4}{v5}
    \pgfnodeconnline{v5}{v6}
    \pgfnodeconnline{v6}{v7}
    \pgfnodeconnline{v7}{v8}
    \pgfnodeconnline{v8}{v9}
    \pgfnodeconnline{v9}{v10}
    \pgfnodeconnline{v10}{v11}
    \pgfnodeconnline{v11}{v12}
    \pgfnodeconnline{v12}{v1}
    \pgfnodeconnline{v6}{v8}
    \pgfnodeconnline{v2}{v4}
    \pgfnodeconnline{v10}{v12}
    \pgfputat{\pgfxy(0+6,4.5+1.5-1+0.5)}{\pgfbox[center,center]{\large $v_1$}}
    \pgfputat{\pgfxy(1+6,4+1.5-1+0.5)}{\pgfbox[center,center]{\large $v_2$}}
    \pgfputat{\pgfxy(2.25+6,3.5+1.5-1+0.5)}{\pgfbox[center,center]{\large $v_3$}}
    \pgfputat{\pgfxy(2.25+6+0.5,2.25+1.5-1)}{\pgfbox[center,center]{\large $v_4$}}
    \pgfputat{\pgfxy(1.75+6+0.5,1.25+1.5-1)}{\pgfbox[center,center]{\large $v_5$}}
    \pgfputat{\pgfxy(1+6+0.5,0.5+1.5-1)}{\pgfbox[center,center]{\large $v_6$}}
    \pgfputat{\pgfxy(0+6+0.5,0+1.5-1)}{\pgfbox[center,center]{\large $v_7$}}
    \pgfputat{\pgfxy(-1+6-0.5,0.5+1.5-1)}{\pgfbox[center,center]{\large $v_8$}}
    \pgfputat{\pgfxy(-1.75+6-0.5,1.25+1.5-1)}{\pgfbox[center,center]{\large $v_9$}}
    \pgfputat{\pgfxy(-2.25+6-0.5,2.25+1.5-1)}{\pgfbox[center,center]{\large $v_{10}$}}
    \pgfputat{\pgfxy(-2.25+6,3.5+1.5-1+0.5)}{\pgfbox[center,center]{\large $v_{11}$}}
    \pgfputat{\pgfxy(-1+6,4+1.5-1+0.5)}{\pgfbox[center,center]{\large $v_{12}$}}

\end{pgfpicture}
\caption{The shield graph}\label{F1}
\end{center}
\end{figure}

Let $G$ be an $r$-regular graph with $n$ vertices. Then the number of its edges is
exactly $\frac{rn}{2}$. Suppose $G$ admits some efficient edge
dominating set $D$. It will be useful to determine $|D|$. Let
$(B=\emptyset,Y,W)$ be the 3-coloring associated to $D$. It is
easy to see, $|Y|=2|D|$, $|W|=n-2|D|$,
$\frac{rn}{2}-|D|=r|W|=r(n-2|D|)=rn-2r|D|$ and
$\frac{rn}{2}-|D|=(r-1)|Y|=(r-1)2|D|=2r|D|-2|D|$ which implies
$rn-2r|D|=2r|D|-2|D|$ and $rn=4r|D|-2|D|=(4r-2)|D|$. Hence,
$|D|=\frac{rn}{4r-2}$.

Now, let us turn to the perfect edge domination problem.
Start by determining the possible perfect edge dominating sets for the
graph of Figure \ref{F1}.

\begin{figure}
\begin{center}
\centering
\begin{pgfpicture}{0cm}{0cm}{12cm}{6cm}
    \pgfsetlinewidth{1pt}

    \pgfnodecircle{v1}[stroke]{\pgfxy(0+6,4.5+1.5-1)}{2mm}
    \pgfnodecircle{v3}[stroke]{\pgfxy(2.25+6,3.5+1.5-1)}{2mm}
    \pgfnodecircle{v5}[stroke]{\pgfxy(1.75+6,1.25+1.5-1)}{2mm}
    \pgfnodecircle{v7}[stroke]{\pgfxy(0+6,0+1.5-1)}{2mm}
    \pgfnodecircle{v9}[stroke]{\pgfxy(-1.75+6,1.25+1.5-1)}{2mm}
    \pgfnodecircle{v11}[stroke]{\pgfxy(-2.25+6,3.5+1.5-1)}{2mm}
    \color{yellow}
    \pgfnodecircle{v2}[fill]{\pgfxy(1+6,4+1.5-1)}{2mm}
    \pgfnodecircle{v4}[fill]{\pgfxy(2.25+6,2.25+1.5-1)}{2mm}
    \pgfnodecircle{v6}[fill]{\pgfxy(1+6,0.5+1.5-1)}{2mm}
    \pgfnodecircle{v8}[fill]{\pgfxy(-1+6,0.5+1.5-1)}{2mm}
    \pgfnodecircle{v10}[fill]{\pgfxy(-2.25+6,2.25+1.5-1)}{2mm}
    \pgfnodecircle{v12}[fill]{\pgfxy(-1+6,4+1.5-1)}{2mm}
    \color{black}
    \pgfnodeconnline{v1}{v2}
    \pgfnodeconnline{v2}{v3}
    \pgfnodeconnline{v3}{v4}
    \pgfnodeconnline{v4}{v5}
    \pgfnodeconnline{v5}{v6}
    \pgfnodeconnline{v6}{v7}
    \pgfnodeconnline{v7}{v8}
    \pgfnodeconnline{v8}{v9}
    \pgfnodeconnline{v9}{v10}
    \pgfnodeconnline{v10}{v11}
    \pgfnodeconnline{v11}{v12}
    \pgfnodeconnline{v12}{v1}
    \pgfsetlinewidth{3pt}
    \pgfnodeconnline{v6}{v8}
    \pgfnodeconnline{v2}{v4}
    \pgfnodeconnline{v10}{v12}
    \pgfputat{\pgfxy(0+6,4.5+1.5-1+0.5)}{\pgfbox[center,center]{\large $v_1$}}
    \pgfputat{\pgfxy(1+6,4+1.5-1+0.5)}{\pgfbox[center,center]{\large $v_2$}}
    \pgfputat{\pgfxy(2.25+6,3.5+1.5-1+0.5)}{\pgfbox[center,center]{\large $v_3$}}
    \pgfputat{\pgfxy(2.25+6+0.5,2.25+1.5-1)}{\pgfbox[center,center]{\large $v_4$}}
    \pgfputat{\pgfxy(1.75+6+0.5,1.25+1.5-1)}{\pgfbox[center,center]{\large $v_5$}}
    \pgfputat{\pgfxy(1+6+0.5,0.5+1.5-1)}{\pgfbox[center,center]{\large $v_6$}}
    \pgfputat{\pgfxy(0+6+0.5,0+1.5-1)}{\pgfbox[center,center]{\large $v_7$}}
    \pgfputat{\pgfxy(-1+6-0.5,0.5+1.5-1)}{\pgfbox[center,center]{\large $v_8$}}
    \pgfputat{\pgfxy(-1.75+6-0.5,1.25+1.5-1)}{\pgfbox[center,center]{\large $v_9$}}
    \pgfputat{\pgfxy(-2.25+6-0.5,2.25+1.5-1)}{\pgfbox[center,center]{\large $v_{10}$}}
    \pgfputat{\pgfxy(-2.25+6,3.5+1.5-1+0.5)}{\pgfbox[center,center]{\large $v_{11}$}}
    \pgfputat{\pgfxy(-1+6,4+1.5-1+0.5)}{\pgfbox[center,center]{\large $v_{12}$}}

\end{pgfpicture}
\caption{The minimum PEDS which is a EEDS}\label{F2}
\end{center}
\end{figure}

\begin{figure}
\begin{center}
\centering
\begin{pgfpicture}{0cm}{0cm}{12cm}{6cm}
    \pgfsetlinewidth{1pt}

    \color{yellow}
    \pgfnodecircle{v11}[fill]{\pgfxy(-2.25+6,3.5+1.5-1)}{2mm}
    \pgfnodecircle{v12}[fill]{\pgfxy(-1+6,4+1.5-1)}{2mm}
    \pgfnodecircle{v2}[fill]{\pgfxy(1+6,4+1.5-1)}{2mm}
    \pgfnodecircle{v9}[fill]{\pgfxy(-1.75+6,1.25+1.5-1)}{2mm}
    \pgfnodecircle{v3}[fill]{\pgfxy(2.25+6,3.5+1.5-1)}{2mm}
    \pgfnodecircle{v5}[fill]{\pgfxy(1.75+6,1.25+1.5-1)}{2mm}
    \color{black}
    \pgfnodecircle{v1}[stroke]{\pgfxy(0+6,4.5+1.5-1)}{2mm}
    \pgfnodecircle{v4}[stroke]{\pgfxy(2.25+6,2.25+1.5-1)}{2mm}
    \pgfnodecircle{v6}[fill]{\pgfxy(1+6,0.5+1.5-1)}{2mm}
    \pgfnodecircle{v7}[fill]{\pgfxy(0+6,0+1.5-1)}{2mm}
    \pgfnodecircle{v8}[fill]{\pgfxy(-1+6,0.5+1.5-1)}{2mm}
    \pgfnodecircle{v10}[stroke]{\pgfxy(-2.25+6,2.25+1.5-1)}{2mm}
    \pgfnodeconnline{v1}{v2}
    \pgfnodeconnline{v3}{v4}
    \pgfnodeconnline{v4}{v5}
    \pgfnodeconnline{v9}{v10}
    \pgfnodeconnline{v10}{v11}
    \pgfnodeconnline{v12}{v1}
    \pgfnodeconnline{v2}{v4}
    \pgfnodeconnline{v10}{v12}
    \pgfsetlinewidth{3pt}
    \pgfnodeconnline{v11}{v12}
    \pgfnodeconnline{v2}{v3}
    \pgfnodeconnline{v5}{v6}
    \pgfnodeconnline{v6}{v7}
    \pgfnodeconnline{v7}{v8}
    \pgfnodeconnline{v8}{v9}
    \pgfnodeconnline{v6}{v8}
    \pgfputat{\pgfxy(0+6,4.5+1.5-1+0.5)}{\pgfbox[center,center]{\large $v_1$}}
    \pgfputat{\pgfxy(1+6,4+1.5-1+0.5)}{\pgfbox[center,center]{\large $v_2$}}
    \pgfputat{\pgfxy(2.25+6,3.5+1.5-1+0.5)}{\pgfbox[center,center]{\large $v_3$}}
    \pgfputat{\pgfxy(2.25+6+0.5,2.25+1.5-1)}{\pgfbox[center,center]{\large $v_4$}}
    \pgfputat{\pgfxy(1.75+6+0.5,1.25+1.5-1)}{\pgfbox[center,center]{\large $v_5$}}
    \pgfputat{\pgfxy(1+6+0.5,0.5+1.5-1)}{\pgfbox[center,center]{\large $v_6$}}
    \pgfputat{\pgfxy(0+6+0.5,0+1.5-1)}{\pgfbox[center,center]{\large $v_7$}}
    \pgfputat{\pgfxy(-1+6-0.5,0.5+1.5-1)}{\pgfbox[center,center]{\large $v_8$}}
    \pgfputat{\pgfxy(-1.75+6-0.5,1.25+1.5-1)}{\pgfbox[center,center]{\large $v_9$}}
    \pgfputat{\pgfxy(-2.25+6-0.5,2.25+1.5-1)}{\pgfbox[center,center]{\large $v_{10}$}}
    \pgfputat{\pgfxy(-2.25+6,3.5+1.5-1+0.5)}{\pgfbox[center,center]{\large $v_{11}$}}
    \pgfputat{\pgfxy(-1+6,4+1.5-1+0.5)}{\pgfbox[center,center]{\large $v_{12}$}}

    \pgfsetlinewidth{1pt}
    \color{yellow}
    \pgfnodecircle{v7}[fill]{\pgfxy(6.3+0+6,0+1.5-1)}{2mm}
    \pgfnodecircle{v8}[fill]{\pgfxy(6.3+-1+6,0.5+1.5-1)}{2mm}
    \pgfnodecircle{v10}[fill]{\pgfxy(6.3+-2.25+6,2.25+1.5-1)}{2mm}
    \pgfnodecircle{v11}[fill]{\pgfxy(6.3+-2.25+6,3.5+1.5-1)}{2mm}
    \pgfnodecircle{v1}[fill]{\pgfxy(6.3+0+6,4.5+1.5-1)}{2mm}
    \pgfnodecircle{v5}[fill]{\pgfxy(6.3+1.75+6,1.25+1.5-1)}{2mm}
    \color{black}
    \pgfnodecircle{v2}[fill]{\pgfxy(6.3+1+6,4+1.5-1)}{2mm}
    \pgfnodecircle{v3}[fill]{\pgfxy(6.3+2.25+6,3.5+1.5-1)}{2mm}
    \pgfnodecircle{v4}[fill]{\pgfxy(6.3+2.25+6,2.25+1.5-1)}{2mm}
    \pgfnodecircle{v6}[stroke]{\pgfxy(6.3+1+6,0.5+1.5-1)}{2mm}
    \pgfnodecircle{v12}[stroke]{\pgfxy(6.3+-1+6,4+1.5-1)}{2mm}
    \pgfnodecircle{v9}[stroke]{\pgfxy(6.3+-1.75+6,1.25+1.5-1)}{2mm}
    \pgfnodeconnline{v9}{v10}
    \pgfnodeconnline{v12}{v1}
    \pgfnodeconnline{v10}{v12}
    \pgfnodeconnline{v11}{v12}
    \pgfnodeconnline{v5}{v6}
    \pgfnodeconnline{v6}{v7}
    \pgfnodeconnline{v8}{v9}
    \pgfnodeconnline{v6}{v8}
    \pgfsetlinewidth{3pt}
    \pgfnodeconnline{v1}{v2}
    \pgfnodeconnline{v2}{v3}
    \pgfnodeconnline{v3}{v4}
    \pgfnodeconnline{v4}{v5}
    \pgfnodeconnline{v2}{v4}
    \pgfnodeconnline{v7}{v8}
    \pgfnodeconnline{v10}{v11}
    \pgfputat{\pgfxy(6.3+0+6,4.5+1.5-1+0.5)}{\pgfbox[center,center]{\large $v_1$}}
    \pgfputat{\pgfxy(6.3+1+6,4+1.5-1+0.5)}{\pgfbox[center,center]{\large $v_2$}}
    \pgfputat{\pgfxy(6.3+2.25+6,3.5+1.5-1+0.5)}{\pgfbox[center,center]{\large $v_3$}}
    \pgfputat{\pgfxy(6.3+2.25+6+0.5,2.25+1.5-1)}{\pgfbox[center,center]{\large $v_4$}}
    \pgfputat{\pgfxy(6.3+1.75+6+0.5,1.25+1.5-1)}{\pgfbox[center,center]{\large $v_5$}}
    \pgfputat{\pgfxy(6.3+1+6+0.5,0.5+1.5-1)}{\pgfbox[center,center]{\large $v_6$}}
    \pgfputat{\pgfxy(6.3+0+6+0.5,0+1.5-1)}{\pgfbox[center,center]{\large $v_7$}}
    \pgfputat{\pgfxy(6.3+-1+6-0.5,0.5+1.5-1)}{\pgfbox[center,center]{\large $v_8$}}
    \pgfputat{\pgfxy(6.3+-1.75+6-0.5,1.25+1.5-1)}{\pgfbox[center,center]{\large $v_9$}}
    \pgfputat{\pgfxy(6.3+-2.25+6-0.5,2.25+1.5-1)}{\pgfbox[center,center]{\large $v_{10}$}}
    \pgfputat{\pgfxy(6.3+-2.25+6,3.5+1.5-1+0.5)}{\pgfbox[center,center]{\large $v_{11}$}}
    \pgfputat{\pgfxy(6.3+-1+6,4+1.5-1+0.5)}{\pgfbox[center,center]{\large $v_{12}$}}

    \pgfsetlinewidth{1pt}
    \color{yellow}
    \pgfnodecircle{v1}[fill]{\pgfxy(-6.3+0+6,4.5+1.5-1)}{2mm}
    \pgfnodecircle{v9}[fill]{\pgfxy(-6.3+-1.75+6,1.25+1.5-1)}{2mm}
    \pgfnodecircle{v3}[fill]{\pgfxy(-6.3+2.25+6,3.5+1.5-1)}{2mm}
    \pgfnodecircle{v4}[fill]{\pgfxy(-6.3+2.25+6,2.25+1.5-1)}{2mm}
    \pgfnodecircle{v6}[fill]{\pgfxy(-6.3+1+6,0.5+1.5-1)}{2mm}
    \pgfnodecircle{v7}[fill]{\pgfxy(-6.3+0+6,0+1.5-1)}{2mm}
    \color{black}
    \pgfnodecircle{v10}[fill]{\pgfxy(-6.3+-2.25+6,2.25+1.5-1)}{2mm}
    \pgfnodecircle{v11}[fill]{\pgfxy(-6.3+-2.25+6,3.5+1.5-1)}{2mm}
    \pgfnodecircle{v12}[fill]{\pgfxy(-6.3+-1+6,4+1.5-1)}{2mm}
    \pgfnodecircle{v2}[stroke]{\pgfxy(-6.3+1+6,4+1.5-1)}{2mm}
    \pgfnodecircle{v8}[stroke]{\pgfxy(-6.3+-1+6,0.5+1.5-1)}{2mm}
    \pgfnodecircle{v5}[stroke]{\pgfxy(-6.3+1.75+6,1.25+1.5-1)}{2mm}
    \pgfnodeconnline{v1}{v2}
    \pgfnodeconnline{v2}{v3}
    \pgfnodeconnline{v4}{v5}
    \pgfnodeconnline{v2}{v4}
    \pgfnodeconnline{v5}{v6}
    \pgfnodeconnline{v7}{v8}
    \pgfnodeconnline{v8}{v9}
    \pgfnodeconnline{v6}{v8}
    \pgfsetlinewidth{3pt}
    \pgfnodeconnline{v3}{v4}
    \pgfnodeconnline{v6}{v7}
    \pgfnodeconnline{v9}{v10}
    \pgfnodeconnline{v10}{v11}
    \pgfnodeconnline{v11}{v12}
    \pgfnodeconnline{v12}{v1}
    \pgfnodeconnline{v10}{v12}
    \pgfputat{\pgfxy(-6.3+0+6,4.5+1.5-1+0.5)}{\pgfbox[center,center]{\large $v_1$}}
    \pgfputat{\pgfxy(-6.3+1+6,4+1.5-1+0.5)}{\pgfbox[center,center]{\large $v_2$}}
    \pgfputat{\pgfxy(-6.3+2.25+6,3.5+1.5-1+0.5)}{\pgfbox[center,center]{\large $v_3$}}
    \pgfputat{\pgfxy(-6.3+2.25+6+0.5,2.25+1.5-1)}{\pgfbox[center,center]{\large $v_4$}}
    \pgfputat{\pgfxy(-6.3+1.75+6+0.5,1.25+1.5-1)}{\pgfbox[center,center]{\large $v_5$}}
    \pgfputat{\pgfxy(-6.3+1+6+0.5,0.5+1.5-1)}{\pgfbox[center,center]{\large $v_6$}}
    \pgfputat{\pgfxy(-6.3+0+6+0.5,0+1.5-1)}{\pgfbox[center,center]{\large $v_7$}}
    \pgfputat{\pgfxy(-6.3+-1+6-0.5,0.5+1.5-1)}{\pgfbox[center,center]{\large $v_8$}}
    \pgfputat{\pgfxy(-6.3+-1.75+6-0.5,1.25+1.5-1)}{\pgfbox[center,center]{\large $v_9$}}
    \pgfputat{\pgfxy(-6.3+-2.25+6-0.5,2.25+1.5-1)}{\pgfbox[center,center]{\large $v_{10}$}}
    \pgfputat{\pgfxy(-6.3+-2.25+6,3.5+1.5-1+0.5)}{\pgfbox[center,center]{\large $v_{11}$}}
    \pgfputat{\pgfxy(-6.3+-1+6,4+1.5-1+0.5)}{\pgfbox[center,center]{\large $v_{12}$}}

\end{pgfpicture}
\caption{Proper PEDS}\label{F3}
\end{center}
\end{figure}

\begin{figure}
\begin{center}
\centering
\begin{pgfpicture}{0cm}{0cm}{12cm}{6cm}
    \pgfsetlinewidth{1pt}

    \pgfnodecircle{v1}[fill]{\pgfxy(0+6,4.5+1.5-1)}{2mm}
    \pgfnodecircle{v3}[fill]{\pgfxy(2.25+6,3.5+1.5-1)}{2mm}
    \pgfnodecircle{v5}[fill]{\pgfxy(1.75+6,1.25+1.5-1)}{2mm}
    \pgfnodecircle{v7}[fill]{\pgfxy(0+6,0+1.5-1)}{2mm}
    \pgfnodecircle{v9}[fill]{\pgfxy(-1.75+6,1.25+1.5-1)}{2mm}
    \pgfnodecircle{v11}[fill]{\pgfxy(-2.25+6,3.5+1.5-1)}{2mm}
    \pgfnodecircle{v2}[fill]{\pgfxy(1+6,4+1.5-1)}{2mm}
    \pgfnodecircle{v4}[fill]{\pgfxy(2.25+6,2.25+1.5-1)}{2mm}
    \pgfnodecircle{v6}[fill]{\pgfxy(1+6,0.5+1.5-1)}{2mm}
    \pgfnodecircle{v8}[fill]{\pgfxy(-1+6,0.5+1.5-1)}{2mm}
    \pgfnodecircle{v10}[fill]{\pgfxy(-2.25+6,2.25+1.5-1)}{2mm}
    \pgfnodecircle{v12}[fill]{\pgfxy(-1+6,4+1.5-1)}{2mm}
    \pgfsetlinewidth{3pt}
    \pgfnodeconnline{v1}{v2}
    \pgfnodeconnline{v2}{v3}
    \pgfnodeconnline{v3}{v4}
    \pgfnodeconnline{v4}{v5}
    \pgfnodeconnline{v5}{v6}
    \pgfnodeconnline{v6}{v7}
    \pgfnodeconnline{v7}{v8}
    \pgfnodeconnline{v8}{v9}
    \pgfnodeconnline{v9}{v10}
    \pgfnodeconnline{v10}{v11}
    \pgfnodeconnline{v11}{v12}
    \pgfnodeconnline{v12}{v1}
    \pgfnodeconnline{v6}{v8}
    \pgfnodeconnline{v2}{v4}
    \pgfnodeconnline{v10}{v12}
    \pgfputat{\pgfxy(0+6,4.5+1.5-1+0.5)}{\pgfbox[center,center]{\large $v_1$}}
    \pgfputat{\pgfxy(1+6,4+1.5-1+0.5)}{\pgfbox[center,center]{\large $v_2$}}
    \pgfputat{\pgfxy(2.25+6,3.5+1.5-1+0.5)}{\pgfbox[center,center]{\large $v_3$}}
    \pgfputat{\pgfxy(2.25+6+0.5,2.25+1.5-1)}{\pgfbox[center,center]{\large $v_4$}}
    \pgfputat{\pgfxy(1.75+6+0.5,1.25+1.5-1)}{\pgfbox[center,center]{\large $v_5$}}
    \pgfputat{\pgfxy(1+6+0.5,0.5+1.5-1)}{\pgfbox[center,center]{\large $v_6$}}
    \pgfputat{\pgfxy(0+6+0.5,0+1.5-1)}{\pgfbox[center,center]{\large $v_7$}}
    \pgfputat{\pgfxy(-1+6-0.5,0.5+1.5-1)}{\pgfbox[center,center]{\large $v_8$}}
    \pgfputat{\pgfxy(-1.75+6-0.5,1.25+1.5-1)}{\pgfbox[center,center]{\large $v_9$}}
    \pgfputat{\pgfxy(-2.25+6-0.5,2.25+1.5-1)}{\pgfbox[center,center]{\large $v_{10}$}}
    \pgfputat{\pgfxy(-2.25+6,3.5+1.5-1+0.5)}{\pgfbox[center,center]{\large $v_{11}$}}
    \pgfputat{\pgfxy(-1+6,4+1.5-1+0.5)}{\pgfbox[center,center]{\large $v_{12}$}}

\end{pgfpicture}
\caption{The trivial PEDS}\label{F4}
\end{center}
\end{figure}

\begin{lemma}\label{T2}
The only perfect edge dominating sets of the shield graph are the
following cases:
\begin{description}
\item[(i)] $\{v_2v_4, v_6v_8, v_{10}v_{12}\}$   (Figure \ref{F2}).
Actually, it is an efficient edge dominating set.
\item[(ii)] $\{v_3v_4, v_6v_7, v_9v_{10}, v_{10}v_{11}, v_{11}v_{12},
v_{12}v_{1},v_{10}v_{12}\}, \{v_2v_3, v_5v_6, v_6v_{7},
v_{7}v_{8}, v_{8}v_{9}, v_{6}v_{8}, v_{11}v_{12}\}$ and $\{v_1v_2,
 v_2v_3,v_3v_{4},$ $v_{4}v_{5}, v_{2}v_{4},
v_{7}v_{8},v_{10}v_{11}\}$ (Figure \ref{F3}). They are proper
perfect edge dominating sets and are symmetrical to each other.
\item[(iii)] The trivial perfect edge dominating set. (Figure
\ref{F4}).
\end{description}
\end{lemma}

\proof Clearly, all edges subsets described in (i), (ii) and (iii)
are perfect edge dominating sets of the shield graph. Let $P$ be
a perfect edge dominating set of such a graph and $(B,Y,W)$, the
3-coloring associated to $P$. Consider the colors of $v_3,v_7$ and
$v_{11}$.
\begin{itemize}
\item The vertices $v_3,v_7,v_{11}\in B$. By Observation
\ref{O5}, $v_2,v_4,v_6,v_8,v_{10}$ and $v_{12}$ are also black
vertices. $v_1, v_5$ and $v_9$ can be neither white vertices 
(Observation \ref{O3}) nor yellow vertices (Observation \ref{O2}).
Hence, all vertices have black colors which implies $P$ to be the
trivial perfect edge dominating set described in (iii). 
\item Some of these vertices are white. Without loss of generality, $v_3$ has
white color. By Observation \ref{O5}, $v_2$ and $v_4$ are yellow
vertices. By observation \ref{O2}, $v_1$ and $v_5$ must be white vertices.
And by Observation \ref{O3}, $v_{12}$ and $v_6$ are
yellow vertices. Clearly, by Observation \ref{O5}, exactly one of
$v_7$ or $v_8$ is yellow vertex and the other is white. Suppose
that $v_7\in Y$ and $v_8 \in W$. Then $v_9\in Y$ (Observation
\ref{O3}) and $v_{10}\notin W$ (Observation \ref{O2}). As $v_{10}$
has two yellow neighbors ($v_9$ and $v_{12}$), by Observation
\ref{O2}, $v_{10}$ is a black vertex. Therefore, the condition of
Observation \ref{O5} does not hold for the triangle
$\{v_{10},v_{11},v_{12}\}$, a contradiction. Consequently, $v_8\in
Y$ and $v_7 \in W$. Symmetrically, using exactly the same
argument, we conclude that $v_{10}\in Y$ and $v_{11} \in W$.
Finally, $v_9$ is a white vertex by Observation \ref{O2} and
clearly $P$ is the efficient edge dominating set described in (i).
\item One of these vertices is yellow. Without loss of generality,
$v_3$ has yellow color. By Observation \ref{O5}, exactly one of
$v_2$ or $v_4$ is a yellow vertex and the other is a white vertex.
Again, without loss of generality $v_2\in Y$ and $v_4\in W$. By
Observation \ref{O3}, $v_5\in Y$ and by Observation \ref{O2}, $v_1
\in W$. By Observation \ref{O3}, $v_{12}$ must be yellow vertex.
Clearly, by Observation \ref{O5}, exactly one of $v_{10}$ or
$v_{11}$ is yellow vertex and the other is white. Suppose that
$v_{10}\in Y$ and $v_{11}\in W$, then $v_9\in W$ (Observation
\ref{O2}) and  $v_8\in Y$ (Observation \ref{O3}). On the other
hand, as $v_5\in Y$, by Observation \ref{O2}, $v_6\notin W$ and
has two yellow neighbors ($v_5$ and $v_8$) which implies by
Observation \ref{O2}, $v_6\in B$. Hence, the condition of
Observation \ref{O5} does not hold for the triangle
$\{v_{6},v_{7},v_{8}\}$ again a contradiction. Consequently,
$v_{11}\in Y$ and $v_{10} \in W$. By Observation \ref{O3}, $v_9\in
Y$ and by Observation \ref{O2}, $v_6, v_8\notin W$. As $v_6$
($v_8$) has two non-white neighbors, its color must be black.
Finally by Observation \ref{O5}, $v_7 \in B$, and clearly $P$ is
the second perfect edge dominating set described in (ii). Making
the other choices of options, leads to the other two perfect edge
dominating sets of (ii). 
$_\triangle$
\end{itemize}

We employ an operation that replaces certain vertices by the
shield graph.

Let $G$ be a graph having a vertex $v$ of degree 3, and $x, y$ and
$z$, the neighbors of $v$.  The {\bf magnification} of $v$ is the
operation that replaces $v$ by a shield $S_v$ and replaces the
edges $xv,yv,zv$ by $xv_3, yv_7,zv_{11}$ (we call $v_3,v_7$ and
$v_{11}$ the {\it contact} vertices of $S_v$). The magnification of a
cubic graph $G$, $\mathcal{M}(G)$, is the graph obtained applying
magnification to all its vertices. The following claim is obvius.
\begin{claim}
$\mathcal{M}(G)$ is claw-free and has degree at most 3. 
\end{claim}
If $G$ has
$n$ vertices (implying that $G$ has $\frac{3n}{2}$ edges) then
$\mathcal{M}(G)$ has $12n$ vertices and
$15n+\frac{3n}{2}=\frac{33n}{2}$ edges.

We relate the cardinality of an efficient edge dominating set of 
$G$ to that of a perfect edge dominating set of $\mathcal{M}(G)$.

\begin{theorem}\label{T3}
Let $G$ be a cubic graph with $n$ vertices and $\mathcal{M}(G)$ the
magnification of $G$. Then $G$ admits some efficient edge
dominating set $D$ if and only if $\mathcal{M}(G)$ has a perfect
edge dominating set $P$ of size at most $\frac{57n}{10}$.
\end{theorem}

\proof Assume $G$ admits some efficient edge dominating set $D$ with
$|D|=\frac{3n}{10}$. Let $Y_D$ the subset of vertices incident to
some edge of $D$ and $W_D=V\setminus Y_D$. It is clear that
$|Y_D|=\frac{3n}{5}$ and $|W_D|=\frac{2n}{5}$. Next, we will
describe how to construct a perfect edge dominating set $P$ of
$\mathcal{M}(G)$ by choosing appropriately its associated
3-coloring $(B,Y,W)$. As the vertices of $\mathcal{M}(G)$ are
partitioned into $n$ different shields $S_v$. we choose locally
the coloring of each shield $S_v$, then we check if the total
coloring is valid. If $v\in W_D$ then its corresponding shield
$S_v$ takes the coloring described in Figure \ref{F2}. Otherwise,
$v\in Y_D$ and $v$ has exactly 2 white neighbors and one neighbor
$v'$ in $G$. In this case, such $S_v$ and $S_{v'}$ should take
some coloring described in Figure \ref{F3} in such a way that the
edge connecting both shields in $\mathcal{M}(G)$ has both extreme
vertices with black color. Clearly, this can be achieved. This
coloring is valid because the partial coloring in each shield is
valid and every edge that connects two shields has both extreme
vertices with black color or one with white color and the other
with yellow color. Now, we examine the number of edges of this
perfect edge dominating set $P$. Every shield $S_v$ contributes 3
edges if $v$ is white and 7 edges if $v$ is yellow. The number of
edges that belong to $P$ and connect two shields is exactly the
number of edges in $D$. Consequently, the total number of edges of
$P$ is
$3|W_D|+7|Y_D|+|D|=\frac{6n}{5}+\frac{21n}{5}+\frac{3n}{10}=\frac{57n}{10}$
as required.

\noindent Conversely, suppose $\mathcal{M}(G)$ has a perfect edge
dominating set $P$ of size at most $\frac{57n}{10}$. Let $(B,Y,W)$
be the associated 3-coloring of $P$. We examine locally the
coloring of each shield $S_v$. If some contact vertex has white
color, then using the same argument as in the proof of Lemma
\ref{T2} all contact vertices have white color and there are
exactly 3 edges of $P$ within $S_v$. In this case,  call $S_v$  a
white shield. Similarly, if some contact vertex has yellow color,
there are exactly 2 contact vertices with yellow color, the other
contact vertex is black and there are exactly 7 edges of $P$
within $S_v$. In this case,  call $S_v$ is a yellow shield. The
third possibility is that all contact vertices have black color
and all 15 edges of $S_v$ belong to $P$. We call $S_v$ a black
shield. Denote by $b,y,w$ the numbers of black, yellow and white
shields, respectively. Clearly, every white shield is adjacent to
exactly 3 yellow shields (two shields are adjacent if there is
some edge between their contact vertices),  and every yellow
shield is adjacent to two white shields and to one non-white shield. It
implies that $3w=2y$. This is the number of edges that connect two
different shields and do not belong to $P$. As the total number of
edges of $\mathcal{M}(G)$ is $\frac{33n}{2}$ and $|P|\leq
\frac{57n}{10}$, the number of edges that do not belong
to $P$ is $12w+8y+3w=12w+12w+3w=27w\geq\frac{108n}{10}$. That is
$w\geq\frac{2n}{5}$. Hence, $y\geq\frac{3n}{5}$ and $b=0$.
Moreover, $w=\frac{2n}{5}$ and $y=\frac{3n}{5}$. We can define the
color of each vertex $v$ of $G$ as the color of $S_v$. This
coloring corresponds to an efficient edge dominating set of $G$.
$_\triangle$

The next corollary is a consequence of Theorems \ref{T1} and
\ref{T3}.

\begin{corollary}\label{NP-claw-free-bounded}
The minimum perfect edge domination problem for claw-free
graphs of degree at most 3 is NP-hard.
\end{corollary}

\proof Clearly, checking if a subset of edges is a perfect edge
dominating set of at most certain size can be done in polynomial
time. This implies the problem is ${\mathcal NP}$. To prove the
completeness, we can apply the reduction from efficient edge
domination problem for cubic graphs which is NP-complete by
Theorem \ref{T1}. Take any instance of this problem. The input is
a cubic graph $G$ with $n$ vertices. We transform $G$ to
$\mathcal{M}(G)$ which is claw-free and has degree at most 3. This
can be done in polynomial-time. By Theorem \ref{T3}, $G$ has an
efficient edge dominating set if only if $\mathcal{M}(G)$ has a
perfect edge dominating set of size at most $\frac{57n}{10}$.
Therefore, the proof is complete.
$_\triangle$

\section{Cubic claw-free graphs}

In this section, we show that the weighted version of the perfect
edge domination problem can be solved in linear time for cubic
claw-free graphs. Without loss of generality, we consider only connected 
graphs.

The first lemma concerns efficient edge dominating sets of
claw-free graphs.

\begin{lemma}\label{DIM-claw-free}\cite{Li-Mi-Sz-14}
There is an algorithm of complexity $O(n)$ which finds the least
weight efficient edge dominating set of a claw-free graph $G$,
or reports that $G$ does not admit efficient edge dominating sets.
\end{lemma}

The following observation is useful.

\begin{observation}\label{vertex-triangle}
A graph $G$ is a cubic claw-free graph if and only if $G$ is cubic
and every vertex of it is contained in some triangle.
\end{observation}

The second lemma concerns proper perfect edge dominating sets of connected graphs where every vertex is contained in some triangle.

\begin{lemma}\label{proper-cubic-claw-free}
Let $G$ be a connected graph where every vertex is contained in some triangle. Then $G$ admits no proper
perfect edge dominating set.
\end{lemma}


\proof Suppose the contrary and let $P$ be a proper perfect edge
dominating set of $G$, and $(B,Y,W)$ its corresponding coloring of the
vertices of $G$. The idea is to show that every vertex of $G$ has black color which implies that $P$ is the trivial perfect edge dominating set, which is a contradiction. Since $P$ is proper, $G$ contains some black vertex $v$. By the hypotesis, $v$ is contained in at least one triangle of $G$. Clearly, any pair of adjacent neighbors of $v$ must have black color by Observation \ref{O5}. We can apply iteratively the same reasoning to each new considered black vertex and we call this procedure as {\it black propagation}. If every vertex of $G$ has been considered, we are done. Suppose there is some unconsidered vertex. As $G$ is connected, there must be an edge $uw$ in $G$ such that $u$ has been considered and $w$ has not. It is clear that $u$ and $w$ do not have a common neighbor. By the hypotesis, there is a triangle $C'$ containing $w$. Since $u$ has black color, by Observation \ref{O3}, $w$ cannot be a white vertex. By Observation \ref{O5}, $w$ must have some neighbor with non-white color in $C'$. But $w$ has already a black neighbor $u$ which is not in $C'$. Therefore, $w$ has black color by Observation \ref{O2}. Again, we can apply the black propagation to $w$. Repeating iteratively the same argument, we conclude that all vertices of $G$ must be black, and therefore $G$ cannot contain a proper edge dominating set. 
$_\triangle$

The algorithm follows immediately from the two above lemmas and the Observation \ref{vertex-triangle}. Let $G$ be a connected cubic claw-free graph, whose edges have been assigned weights, possibly negative. By Lemma \ref{proper-cubic-claw-free} it does not contain proper perfect edge dominating sets. Then first,
we apply the algorithm described in \cite{Li-Mi-Sz-14} for claw-free graphs. If $G$ contains an
efficient edge dominating set then the latter algorithm finds the
least weighted of such sets, and the minimum between this set and the trivial perfect edge dominating set $E(G)$ is the answer of our
algorithm. Otherwise, $G$ does not contain an efficient edge
dominating set,  and therefore
the only perfect edge dominating set of $G$ is $E(G)$ and this is
the answer of the algorithm.

By recalling that $|E(G)| = \frac{3 n}{2}$, the complexity of the algorithm is therefore $O(n)$.

This linear time algorithm can be extended easily to claw-free graphs where every degree two vertex has two adjacent neighbors. Because in this case, every vertex is in some triangle except for pendant vertices. We can see that Lemma \ref{proper-cubic-claw-free} is still valid for any connected component of these graphs. We know that pendant vertices never have black colors by Observation \ref{O2}. If there is some black vertex in the connected component, then all vertices of this component are colored black, except the pendant vertices. Therefore, every such pendant vertex has a black neighbor if the component is not exactly $K_2$. In any case, every pendant vertex must be colored yellow. Consequently, there are no white vertices and the lemma is true.  

\section{Bounded degree graphs of large girth}

In this section, we prove NP-hardness of the perfect edge
domination problem in bounded degree graphs even restricted to those of large girth.

The following additional notation is employed. For a non negative
integer $k$, and an edge $e=vw \in E(G)$, where $v,w \in V(G)$,
the \emph{$k$-subdivision} of $e$ is the operation that replaces
$e$ by a $(k+1)$-path, denoted by $S_k(e)$, formed by edges $e_0,
\ldots, e_k$, such that all its $k$ internal vertices are newly
inserted vertices, each of degree 2, while $e_0$ is incident to
$v$ and $e_k$ incident to $w$.  In general, for $E' \subseteq
E$, $S_k(E')$ is the set of the $k$-subdivisions of the edges of
$E'$, while $S_k(G)$ denotes the \emph{$k$-subdivision graph} of
$G$, the one obtained by subdividing all its edges. In such a
graph, the extreme vertices of each edge $e \in E(G)$ become the
extremes of the paths $S_k(e)$ and are called {\it extremes} of
$S_k(G)$. We omit the subscript from $S$ when there is no
ambiguity.

Throughout this section, $G$ denotes an $r$-regular graph, and
$S(G)$ the $3k$-subdivision of $G$, for some integer $k \geq 0$.

\begin{lemma}\label{lem:edge-bound}
Let $e=vw \in E(G)$ and $P$ be a perfect
edge dominating set of $S (G)$. Then
\begin{enumerate}
\item $|S(e) \cap P| \geq k$. 
\item $|S(e) \cap P| = k \Rightarrow$ $e_0,e_{3k} \not \in P$, 
$|\{v,w\}\cap W|=1$ and $|\{v,w\}\cap Y|=1$.
\end{enumerate}
\end{lemma}

\proof
Note that Lemma \ref{lem:edge-bound} holds for $k=0$ since (1.) is trivially true and (2.) holds because $e=e_0=e_k$ is dominated by $P$. 
For $k\geq1$ since $P$ is a perfect edge dominating set, at least
one in every three consecutive edges of $S(e)$  must belong to
$P$. Observe the following subsets of edges
$\{e_1,e_2,e_3\},\{e_4,e_5,e_6\},\dots,
\{e_{3k-2},e_{3k-1},e_{3k}\}$. Each of them has at least one edge
of $P$ and we can conclude that $|S(e) \cap P| \geq k$. If some of
them has at least two edges of $P$ then $|S(e) \cap P| \geq k +
1$. Otherwise, every one has exactly one edge of $P$. Easily, we
can determine these edges knowing which edge of $\{e_1,e_2,e_3\}$
belongs to $P$. The three possibilities are: (i)
$e_1,e_4,\dots,e_{3k-2}$ (ii) $e_2,e_5,\dots,e_{3k-1}$ and (iii)
$e_3,e_6,\dots,e_{3k}$.

Next, we assume that $|S(e) \cap P| = k $ and we consider the
alternatives:
\begin{enumerate}
\item $e_0 \in P$ \\
Then $|S(e) \cap P|=|\{e_0\}|+|\{e_1,e_2,e_3\}\cap
P|+\dots+|\{e_{3k-2},e_{3k-1},e_{3k}\}\cap P|\geq 1+k$ which is a
contradiction.

\item $e_{3k} \in P$ \\
Using similar argument of Alternative 1, we can conclude that
$|S(e) \cap P| \geq k+1$. Again, a contradiction.

\item $e_0, e_{3k} \not \in P$ \\
Examine the further alternatives:
\begin{enumerate}
\item $v,w \in W$: \\
Clearly, $e_1, e_{3k-1} \in P$. Hence, some of the subsets
$\{e_1,e_2,e_3\}$,
$\{e_4,e_5,e_6\}$,... ,$\{e_{3k-2},e_{3k-1},e_{3k}\}$ has 
at least two edges of $P$ and
implies that $|S(e) \cap P| \geq k + 1$. This is a contradiction.

\item $v,w \in Y$: \\
Then $e_1, e_{3k-1} \not \in P$ because $e_0$ ($e_{3k}$) is
dominated by some edge incident to $v$ ($w$). Furthermore, $e_2,
e_{3k-2} \in P$ (to dominate $e_1$ and $e_{3k-1}$, respectively).
Again, some of the subsets $\{e_1,e_2,e_3\},$
$\{e_4,e_5,e_6\}$,... ,$\{e_{3k-2},e_{3k-1},e_{3k}\}$ has 
at least two edges of $P$
implying that $|S(e) \cap P| \geq k + 1$ and leading to a
contradiction.

\item $v \in B$ or $w \in B$: \\
Then $e_0\in P$
 or  $e_{3k} \in P$ which is a contradiction to the
assumption.

\item One of $v,w$ is $W$ and the other one is $Y$. This is the
only situation where $|S(e) \cap P| = k$ can
occur.
\end{enumerate}
\end{enumerate} $_\triangle$

The next lemma describes a lower bound for the size of a perfect
edge dominating set of $S(G)$.

\begin{lemma}\label{lem:general-bound}
Let $e \in E(G)$ and $P$ be a perfect edge dominating set of $S(G)$.
Then
\begin{enumerate}
\item $|S(N'[e]) \cap P| \geq 2rk - k +1$ \item $|P| \geq
\frac{nr}{2(2r-1)}\cdot(2rk - k + 1)$
\end{enumerate}
\end{lemma}

\proof
By Lemma \ref{lem:edge-bound}, it follows that $|S(e)
\cap P| \geq k$, for any $e \in E(G)$. Since $G$ ia an $r$-regular
graph, $|N'[e]| = 2r-1$. Consequently, $|S(N'[e])
 \cap P| \geq (2r-1)k$. In addition, also by Lemma
 \ref{lem:edge-bound}, $|S(e)\cap P| = k$ implies that an extreme
 vertex $v$ of $e$ belongs to $W$, the other extreme $w \in Y$,  and the extreme edge
 $e_{3k}$ of $S(e)$, incident to $w$, is not in $P$. In order to
 attain $|S(N'[e])\cap P| = (2r-1)k$, all $2r-1$ edges of $N'[e]$ must
 satisfy the latter condition. However, the yellow extreme vertex
 of $e$ must have some incident edge $e''\in P$ and this edge $e''$ belongs to some $S(e')$, $e'\in E(G)$. Clearly, $e'\in N'(e)$ and
according to Lemma \ref{lem:edge-bound}, $|S(e') \cap P| \geq k +
1$. Consequently, the bound is refined as $S(N'[e] \cap P) \geq
(2r-2)k + k +1 =  2rk - k +1$, proving the first part of the
lemma.

To prove the second part, we remind that in each edge neighborhood
$N'[e]$ of $G$, $2r-2$ of the edges of $S(N'[e])$ contain at least
$k$ edges of $P$, while one edge of $N'[e]$ corresponds in
$S(N'[e])$ to at least $k+1$ edges in $P$. Since $|E(G)|=
\frac{nr}{2}$ and $|P|=\sum_{e\in E(G)}|S(e)\cap P|$, it follows
$(2r-1)|P| = \sum_{e\in E(G)}|S(N'[e])\cap P| \geq
\frac{nr}{2}(2rk - k + 1)$. That is, $|P| \geq
\frac{nr}{2(2r-1)}\cdot(2rk - k + 1)$, as required.
$_\triangle$

\begin{corollary}\label{cor:coloring}
The following are equivalent:
\begin{description}
\item[(i)] $S(G)$ admits a perfect edge dominating set of size
$\frac{nr}{2(2r-1)}\cdot(2rk - k + 1)$ \item[(ii)]$G$ can be
colored with two colors, $W$ and $Y$, such that
\begin{itemize}
\item No two vertices of $W$ are adjacent, and \item Every vertex
of $Y$ has exactly one neighbor of the same color in $G$.
\end{itemize}
\item[(iii)] $G$ admits an efficient edge dominating set.
\end{description}
\end{corollary}

\proof
(i) $\Rightarrow$ (ii) \\
By the two previous lemmas, whenever $|P| =
\frac{nr}{2(2r-1)}\cdot(2rk - k + 1)$, in each edge neighborhood
$N'[e]$ of $G$, $2r-2$ of the edges of $S(N'[e])$ contain exactly
$k$ edges of $P$, while one edge of $N'[e]$ contributes with $k+1$
edges to $P$. Color the vertices of $G$, as follows. Assign color
$Y$ to both extremes of the edges which contribute with $k+1$
edges and the color $W$ to the remaining vertices of $G$. The
coloring so obtained satisfies the required conditions.

(ii) $\Rightarrow$ (i) \\
By hypothesis, $G$ admits a coloring with colors $W$ and $Y$,
satisfying the conditions of the theorem. Such a coloring
partitions the edges of $E(G)$ into two types. Those whose
extremes are colored $W$ and $Y$, and those with both extremes colored 
$Y$. If $k=0$, $G=S(G)$ and we define $P$ as the set of edges with both extremes colored $Y$. 
If $k\geq1$ since every yellow vertex has exactly one yellow neighbor in
$G$, it follows that, for any $e \in E(G)$, $S(N'[e])$ contains
exactly one edge with both extremes $Y$, while each one of the
remaining edges of $S(N'[e])$ has extremes colored $W$ and $Y$,
respectively. We can construct a perfect edge dominating set $P$,
by selecting edges from each $N'(e)$, where $e=vw \in E(G)$. If
$v$ and $w$ are both in $Y$  then $S(e)\cap
P=\{e_0,e_3,\dots,e_{3k}\}$. Otherwise, we can suppose that $v$ is
in $W$ while $w$ is in $Y$,  and in this case $S(e)\cap
P=\{e_1,e_4,\dots,e_{3k-2}\}$. It is easy to verify that $P$ 
is a perfect edge dominating set of size
$\frac{nr}{2(2r-1)}\cdot(2rk - k + 1)$.

The proofs (ii) $\Rightarrow$ (iii) and (iii) $\Rightarrow$ (ii)
are straightforward.
$_\triangle$

Next, we present the main result of this section.

\begin{theorem}\label{thm:NP-bounded-no-short}
The cardinality perfect edge domination problem is NP-hard,
even if restricted to graphs having vertices of bounded degree
$r$ and girth at least $k$, for any fixed $r,k \geq 3$.
\end{theorem}

\proof It is straightforward to conclude that the problem
belongs to ${\mathcal NP}$. The transformation is from the
efficient edge domination problem for $r$-regular graphs,
which is known to be NP-complete for fixed $r$ (Cardoso, Cerdeira,
Delorme and Silva \cite{Ca-Ce-De-Si}). Let $G$ be an $r$-regular
graph, $|V(G)| = n$. Construct an instance of the perfect edge
domination problem, as follows. Let $k \geq 3$ be a fixed integer,
define $k' = max\{1,\lceil \frac{k-3}{9} \rceil\}$. The input
graph for the latter problem is the $3k'$-subdivision $S(G)$ of
$G$. Set the value of $p$, as $p = \frac{nr}{2(2r-1)}\cdot(2rk' -
k' + 1)$. Clearly, $S(G)$ has no cycles shorter than $3\cdot
(3k'+1)=9k'+3\geq k$ and its vertices have maximum degree $r$.
Finally, by Corollary \ref{cor:coloring}, $G$ contains an
efficient edge dominating set if and only if $S(G)$ contains a
perfect edge dominating set of size $p$ which completes the proof
of NP-completeness.
$_\triangle$

As Lemma \ref{lem:edge-bound}, Lemma \ref{lem:general-bound} and
Corollary \ref{cor:coloring} hold for $k=0$, and using similar
arguments of the proof of Theorem \ref{thm:NP-bounded-no-short},
we can prove the following corollary.

\begin{corollary}\label{thm:NP-regular}
The cardinality perfect edge domination problem is NP-hard,
even if restricted to $r$-regular graphs, for $r \geq 3$.
\end{corollary}

\section{A dichotomy theorem}

In this section, we describe a dichotomy theorem for the
complexity of the perfect edge domination problem. It is a
consequence of the NP-completeness proofs of the previous sections.

Define a {\it linear forest} as a graph whose connected components are induced paths.

\begin{theorem}\label{dichotomy}
Let $H$ be a graph, and ${\mathcal G}$ the class of
$H$-free graphs of degree at most $d$, for some fixed $d \geq 3$. Then the
perfect edge domination problem is
\begin{itemize}
\item polynomial time solvable for graphs in ${\mathcal G}$ if $H$ is a linear forest
\item NP-complete otherwise.
\end{itemize}
\end{theorem}

\proof
First, assume $H$ is not a linear forest. There are
two situations. If $H$ contains some induced cycle $C_s$ it
follows that ${\mathcal G}$ contains, as a subclass, the
$C_s$-free graphs of maximum degree at most $d$. The latter
subclass contains the graphs of bounded degree at most $d$ 
and girth at least $s+1$. By Theorem 
\ref{thm:NP-bounded-no-short}, the perfect edge dominating set is
NP-complete for such a class. For the second alternative, assume
that $H$ does not contain cycles. Since $H$ is not a linear
forest, it follows that $H$ is a forest containing a vertex of
degree at least 3. That is, ${\mathcal G}$ contains the class of
claw-free graphs of maximum degree at most 3. By Corollary
\ref{NP-claw-free-bounded} the perfect edge domination problem is also
NP-complete in this case.

Finally, assume that $H$ is a linear forest and its connected components consist of exactly
 $t$ disjoint induced paths. In this case, we can bound the number
of vertices of each connected component of $G$.
Then $H$ is an induced subgraph of $P_q$, with $q=|V(H)|+t-1$. An
$H$-free graph is in particular $P_q$-free, and each connected
component of $G$ is $P_q$-free and has maximum degree at most $d$.
Then each connected component has at most $\frac{d^{q-1}-1}{d-1}$
vertices. That is, a constant number of vertices as $H$ has fixed
size. Therefore, we can solve perfect edge domination problem
in polynomial time.
$_\triangle$

\begin{corollary}
The perfect edge domination problem is NP-complete for $H$-free graphs where $H$ is any graph except linear forests.
\end{corollary}

We leave as an open problem the question whether or not there exists some graph class for which the efficient edge domination problem is NP-complete and the perfect edge domination problem can be solved in polynomial time.

\section{$P_k$-free graphs and $P_5$-free graphs}

In this section, we describe a robust linear time algorithm for finding
minimum weight perfect edge dominating set of a $P_5$-free
graph. Given an arbitrary graph $G$, in linear time the algorithm
either finds such edge dominating set or exhibits an induced $P_5$
of the graph.

The following definitions are useful. Let $G$ be a connected
graph, and $v \in V(G)$. The {\it eccentricity} of $v$, denoted
$\epsilon(v)$ is the maximum distance between $v$ and any other vertex. A
vertex of minimum eccentricity in the graph is called {\it
central}. Finally, a vertex having eccentricity at most 2 is named
{\it principal vertex}.

Note that if a connected graph has a vertex with eccentricity at least $4$ then it has an induced $P_5$. Any central vertex of a connected $P_5$-free
graph has eccentricity at most $2$ (consequence of Theorem \ref{TTuza}).

\subsection{The basis}

We describe below the theorems in which the correctness and
complexity of the proposed algorithm is based on.

The following structural result by Bacs\'o and Tuza is fundamental.

\begin{theorem}\label{TTuza}
\cite{Ba-Tu} Every connected graph contains an induced $P_5$, a
dominating $K_p$, or a dominating $P_3$.
\end{theorem}

The proposed algorithm needs to determine which of these three
subgraphs $G$ contains. This will be achieved through a principal
vertex. Then we need a robust method to compute such a vertex, if
it exists. For a chosen vertex $v$, call {\it Test(v)} the
operation that  determines its eccentricity and recognizes the
following situations:

\begin{enumerate}
\renewcommand{\labelenumi}{(\roman{enumi})}
\item  If $v$ has infinite eccentricity, then $G$ is not
connected. \item  If $v$ has eccentricity at least $4$, then $G$
has an induced $P_5$.  \item  If $v$ has eccentricity $3$, then
$G$ has an induced $P_4$ starting in $v$, \item  If $v$ has
eccentricity at most $2$, then $v$ is a principal vertex of $G$.
\end{enumerate}

It is clear that Test$(v)$ can be done in linear-time. The cases
(i) and (ii) allow to recognize that the input $G$ is not a
connected $P_5$-free graph.

The next theorem describes the robust linear time computation of a
principal vertex.

\begin{theorem}\label{principal}
Any connected graph $G$ contains a principal vertex or an induced $P_5$.
Moreover there is a linear-time algorithm to find a principal
vertex of $G$ or to detect that $G$ is not $P_5$-free.
\end{theorem}

\proof
We describe an algorithm that uses at most 3 times the procedure
{\it Test} a vertex. At any point of the algorithm,  if an
induced $P_5$ is found, then the algorithm ends and returns that
induced $P_5$. First choose any vertex $u\in V$ and Test$(u)$. If
$\epsilon(u)=3$, let $z$ be such that $dist(u,z)=3$ and $u,v,w,z$
an induced $P_4$. Choose $v_1\in A=N(u)\cap N(w)$ such that $v_1$
has the maximum number of neighbors 
in $B=V\setminus (N(u)\cup N(w)\cup N(z))$.
Note that $u,v_1,w,z$ is an induced $P_4$ and possibly $v=v_1$.

Next, Test$(v_1)$ and Test$(w)$. We will show that
$\epsilon(v_1)=\epsilon(w)=3$ implies that $G$ has an induced
$P_5$. As $v_1$ has eccentricity $3$, there exists a vertex $x$
such that $dist(x,v_1)=3$. Also $x$ verifies $dist(x,w)=3$ or
$dist(x,w)=2$ because $v_1w\in E$.

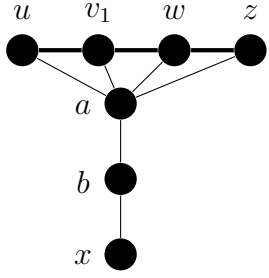
\begin{figure}
\vspace{0.5cm}
\center \tikzstyle{every node}=[circle, fill, inner sep=1.5mm,
label distance=-1mm,node distance=1cm, font=\large]
\begin{tikzpicture}[]
  \node (u) []  {};
  \node (v1) [right of=u] {};
  \node (w) [right of=v1] {};
  \node (z) [right of=w] {};
  \node (a) [below left of=w] {};
  \node (b) [below of=a] {};
  \node (x) [below of=b] {};
  \path
    (u) edge [ultra thick] (v1)
            edge (a)
    (v1) edge [ultra thick] (w)
             edge (a)
    (w) edge [ultra thick] (z)
           edge (a)
    (a) edge (b)
   edge (z)
    (b) edge (x);
    \pgfputat{\pgfxy(1,0.5)}{\pgfbox[center,center]{\large $v_1$}}
    \pgfputat{\pgfxy(0,0.5)}{\pgfbox[center,center]{\large $u$}}
    \pgfputat{\pgfxy(2,0.5)}{\pgfbox[center,center]{\large $w$}}
    \pgfputat{\pgfxy(3,0.5)}{\pgfbox[center,center]{\large $z$}}
    \pgfputat{\pgfxy(0.8,-0.75)}{\pgfbox[center,center]{\large $a$}}
    \pgfputat{\pgfxy(0.8,-1.75)}{\pgfbox[center,center]{\large $b$}}
    \pgfputat{\pgfxy(0.8,-2.75)}{\pgfbox[center,center]{\large $x$}}
\end{tikzpicture}
\caption{$dist(v_1,x)=dist(w,x)=3$.} \label{dist1}
\end{figure}

If $dist(x,w)=3$ let $v_1,a,b,x$ be a shortest path between $v_1$
and $x$, as in Figure \ref{dist1}. Note that
$N[x]\cap\{u,v_1,w,z\}=\emptyset$, $N(b)\cap\{v_1,w\}=\emptyset$
because $dist(x,v_1)=3$ and $dist(x,w)=3$. If
$N(b)\cap\{u,z\}\ne\emptyset$ there is an induced $P_5$ and
we can find it and return it. Otherwise,
$N(b)\cap\{u,z\}=\emptyset$. We can assume $wa\in E$ otherwise
$w,v_1,a,b,x$ is an induced $P_5$. Following the same idea,
$x,b,a,v_1,u$ and $x,b,a,w,z$ are possible induced $P_5$'s, then
we can assume that $ua,az\in E$ which is a contradiction because
$dist(u,z)=3$. Therefore, in this case the algorithm always finds an
induced $P_5$.

\begin{figure}
\center \tikzstyle{every node}=[circle, fill, inner
sep=1.5mm,label distance=-1mm,node distance=1cm, font=\large]%
\begin{tikzpicture}[]
  \node (p)  [fill,white] {}; 
  \node (u)  [below of=p] {};
  \node (v1) [right of=u] {};
  \node (a) [below of=v1] {};
  \node (w) [right of=v1] {};
  \node (z) [right of=w] {};
  \node (x) [below of=a] {};
 \node (y) [above of=v1] {};
 \path
    (u) edge [ultra thick, green] (v1)
            edge [ultra thick, green] (a)
    (v1) edge (w)
             edge [ultra thick, green] (y)
    (a) edge [ultra thick] (w)
             edge [ultra thick, green] (x)
             edge [thick] (x)
    (w) edge [ultra thick] (z)
    (x) edge [bend left, dashed, ultra thick] (y);
    \pgfputat{\pgfxy(1+0.4,1.5-1.4)}{\pgfbox[center,center]{\large $y$}}
    \pgfputat{\pgfxy(1.3,0.5-1.15)}{\pgfbox[center,center]{\large $v_1$}}
    \pgfputat{\pgfxy(0+0.2,0.5-1.15)}{\pgfbox[center,center]{\large $u$}}
    \pgfputat{\pgfxy(2+0.2,0.5-1.15)}{\pgfbox[center,center]{\large $w$}}
    \pgfputat{\pgfxy(3+0.2,0.5-1.15)}{\pgfbox[center,center]{\large $z$}}
    \pgfputat{\pgfxy(1.5,-1-1.1)}{\pgfbox[center,center]{\large $a$}}
    \pgfputat{\pgfxy(1.5,-2-1.1)}{\pgfbox[center,center]{\large $x$}}
\end{tikzpicture}
\caption{$dist(v_1,x)=3$ and $dist(w,x)=2$.} \label{dist2}
\end{figure}
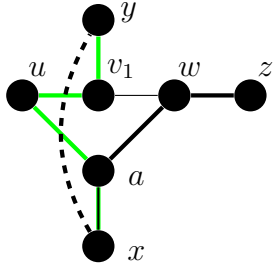

If $dist(x,w)=2$ let $w,a,x$ be a shortest path between $w$ and
$x$. Note that $N[x]\cap\{u,v_1,w\}=\emptyset$ because
$dist(x,v_1)=3$. If $zx\in E$ then  $u,v_1,w,z,x$ is an induced
$P_5$. Thus, $zx\notin E$ and we are in the situation of Figure
\ref{dist2}. We assume $ua\in E$ otherwise $u,v_1,w,a,x$ is an
induced $P_5$. This implies that $a\in A=N(u)\cap N(w)$ and it is
adjacent to $x\in B=V\setminus (N(u)\cup N(w)\cup N(z))$. Recall
that $v_1$ was chosen as a vertex belonging to $A$ with maximum
degree in $B$. As $v_1x\notin E$, then there exists $y\in B$ such
that $v_1y\in E$ and $ay\notin E$. It follows that there is a $P_5$
induced either by $y,v_1,u,a,x$ if $xy\notin E$ or by $x,y,v_1,w,z$ if
$xy\in E$. Again, the algorithm always finds an induced
$P_5$.$_\triangle$

Once a principal vertex is obtained, we find an induced $P_5$, or
a dominating $K_p$, or a dominating $P_3$ following the theorem
below. We remark that using a recent characterization of $P_k$-free
graphs, by Camby and Schaudt \cite{Ca-Sc}, it is possible to obtain a dominating induced $P_3$ or a dominating $K_p$ in
$O(n^5(n+m))$ time when a connected $P_5$-free graph is given.

\begin{theorem}\label{P3orKp}
For any connected graph $G$, there is a linear-time algorithm to
find a dominating induced $P_3$, a dominating $K_p$, or to detect
that $G$ is not a $P_5$-free graph.
\end{theorem}

\begin{figure}
\vspace{0.5cm}
\center \tikzstyle{every node}=[circle, fill, inner sep=1.5mm, node
distance=1cm, label distance=-2mm, font=\large]
\begin{tikzpicture}[]

  \node (v)  [] {};
  \node (v2) [below of=v] {};
  \node (v1) [left of=v2] {};
  \node (v3) [right of=v2] {};
  \node (w1) [below of=v1] {};
  \node (w2) [below of=v2] {};
  \node (w3) [below of=v3] {};
  \node (i) [left of=w1, fill=white] {(i)};
  \path
    (v) edge [ultra thick] (v1)
            edge (v2)
            edge [ultra thick] (v3)
    (v1) edge [ultra thick] (w1)
    (v2) edge (w2)
    (v3) edge (w3)
    (w1) edge [ultra thick] (w2)
             edge [bend right] (w3)
    (w2) edge (w3);

  \node (ii) [node distance=1.8cm, right of=w3, fill=white] {(ii)};
  \node (w'1) [right of=ii] {};
  \node (w'2) [right of=w'1] {};
  \node (w'3) [right of=w'2] {};
  \node (v'1) [above of=w'1] {};
  \node (v'2) [above of=w'2] {};
  \node (v'3) [above of=w'3] {};
  \node (v')  [above of=v'2] {};
  \path
    (v') edge [ultra thick] (v'1)
            edge (v'2)
            edge [ultra thick] (v'3)
    (v'1) edge [ultra thick] (w'1)
    (v'2) edge (w'2)
            edge (v'3)
    (v'3) edge (w'3)
    (w'1) edge [ultra thick] (w'2)
              edge [bend right] (w'3);

  \node (iii) [node distance=1.8cm, right of=w'3, fill=white] {(iii)};
  \node (w''1) [right of=iii] {};
  \node (w''2) [right of=w''1] {};
  \node (w''3) [right of=w''2] {};
  \node (v''1) [above of=w''1] {};
  \node (v''2) [above of=w''2] {};
  \node (v''3) [above of=w''3] {};
  \node (v'')  [above of=v''2] {};
  \path
    (v'') edge [ultra thick] (v''1)
            edge (v''2)
            edge [ultra thick] (v''3)
    (v''1) edge [ultra thick] (w''1)
                 edge (v''2)
    (v''2) edge [ultra thick,green] (w''2)
                 edge (v''3)
    (v''3) edge [ultra thick,green] (w''3)
                 edge [ultra thick,green] (v''2)
    (w''1) edge [dashed,ultra thick] (w''2)
             edge [bend right,ultra thick,green] (w''3);
    \pgfputat{\pgfxy(0,0.5)}{\pgfbox[center,center]{\large $v$}}
    \pgfputat{\pgfxy(1.5-2.95,-1+0.2)}{\pgfbox[center,center]{\large $v_1$}}
    \pgfputat{\pgfxy(1.5-2.95,-2+0.3)}{\pgfbox[center,center]{\large $w_1$}}
    \pgfputat{\pgfxy(1.5-1.1,-1+0.2)}{\pgfbox[center,center]{\large $v_2$}}
    \pgfputat{\pgfxy(1.5-1.1,-2+0.2)}{\pgfbox[center,center]{\large $w_2$}}
    \pgfputat{\pgfxy(1.5,-1+0.2)}{\pgfbox[center,center]{\large $v_3$}}
    \pgfputat{\pgfxy(1.5,-2+0.2)}{\pgfbox[center,center]{\large $w_3$}}
    \pgfputat{\pgfxy(4.8,0.5)}{\pgfbox[center,center]{\large $v$}}
    \pgfputat{\pgfxy(1.5-2.95+4.8,-1+0.2)}{\pgfbox[center,center]{\large $v_1$}}
    \pgfputat{\pgfxy(1.5-2.95+4.8,-2+0.3)}{\pgfbox[center,center]{\large $w_1$}}
    \pgfputat{\pgfxy(1.5-1.1+4.8,-1+0.2)}{\pgfbox[center,center]{\large $v_2$}}
    \pgfputat{\pgfxy(1.5-1.1+4.8,-2+0.2)}{\pgfbox[center,center]{\large $w_2$}}
    \pgfputat{\pgfxy(1.5+4.8,-1+0.2)}{\pgfbox[center,center]{\large $v_3$}}
    \pgfputat{\pgfxy(1.5+4.8,-2+0.2)}{\pgfbox[center,center]{\large $w_3$}}
    \pgfputat{\pgfxy(9.6,0.5)}{\pgfbox[center,center]{\large $v$}}
    \pgfputat{\pgfxy(1.5-2.95+9.6,-1+0.2)}{\pgfbox[center,center]{\large $v_1$}}
    \pgfputat{\pgfxy(1.5-2.95+9.6,-2+0.3)}{\pgfbox[center,center]{\large $w_1$}}
    \pgfputat{\pgfxy(1.5-1.1+9.6,-1+0.2)}{\pgfbox[center,center]{\large $v_2$}}
    \pgfputat{\pgfxy(1.5-1.1+9.6,-2+0.2)}{\pgfbox[center,center]{\large $w_2$}}
    \pgfputat{\pgfxy(1.5+9.6,-1+0.2)}{\pgfbox[center,center]{\large $v_3$}}
    \pgfputat{\pgfxy(1.5+9.6,-2+0.2)}{\pgfbox[center,center]{\large $w_3$}}
\end{tikzpicture}
\caption{Center $v$ and its sons and proper grandsons.}
\label{P3orKpfig}
\end{figure}
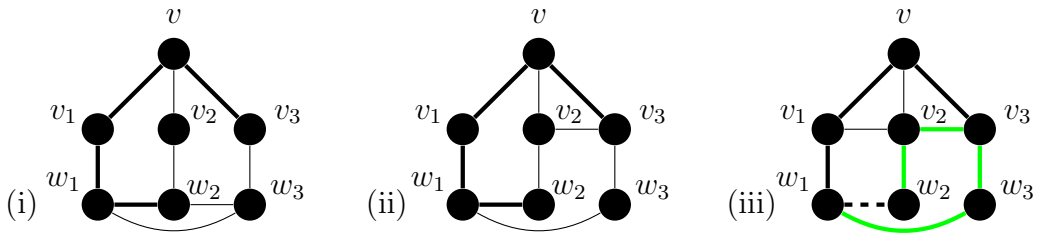

\proof
The algorithm searches a dominating set contained in $N[v]$ where
$v$ is a principal vertex of $G$. Again, at any point of the
algorithm,  if an induced $P_5$ is found, then the algorithm ends
and returns that induced $P_5$.
\begin{enumerate}
\item Find a principal vertex $v$ or an induced $P_5$ of $G$ using
the robust linear-time algorithm of Theorem \ref{principal}. 
\item Let $X:=N[v]$ be the initial dominating set. Consider iteratively 
each vertex $w \in N(v)$: if $X \setminus \{w\}$ is still a 
dominating set then $X:=X\setminus \{w\}$. This can be done in
linear-time using a variable for each vertex of $V\setminus N[v]$
to count the number of neighbors in $X$. A vertex $w\in N(v)$ can
not be removed from $X$ if only if some of its neighbors has
exactly one dominator in $X$. \item If $|X|\leq3$ then $G$ has a
dominating $P_3$ or a dominating $K_{|X|}$. If $|X|\geq4$ and it is
not a complete graph, we will show that $G$ has an induced $P_5$. We
can assume that $G$ has a subgraph like one in Figure
\ref{P3orKpfig}, where $\{v_1,v_2,v_3\}\subset X$ and does not
induce a triangle. Each one of them has a proper dominated vertex
$w_1, w_2, w_3$ respectively. Edges between $w_i$ and $w_j$ are
drawn in order to avoid the induced $P_5$ $w_i,v_i,v,v_j,w_j$.
Nevertheless in (i) and (ii) $w_2,w_1,v_1,v,w_3$ is a induced
$P_5$; and in (iii) the vertices $w_2,v_2,v_3,w_3,w_1$ or
$w_2,w_1,v_1,v,w_3$ induce a $P_5$.
\end{enumerate}$_\triangle$\\

We remark that the above theorem might be of interest to other
$P_5$-free algorithmic problems, since it represents an
algorithmic proof of Bacs\'o and Tuza's theorem \cite{Ba-Tu}.

\subsection{Colorings, vertex dominating sets and perfect edge dominating
sets}

Next, we describe the remaining algorithms on which the proposed
solution is based. They are related to 3-colorings of the graph,
vertex dominating sets and the different kinds of perfect edge
dominating sets, that is, efficient edge dominating, trivial ones
and proper.

The first two theorems concern the construction of perfect edge
dominating sets, associated to a given 3-coloring of the graph.

\begin{theorem}\label{algo}
Given a graph $G$ and $W\subseteq V$, there is a linear-time
algorithm to verify if there exists some perfect edge dominating set
whose associated 3-coloring is $(B,Y,W)$.
\end{theorem}

\proof In the affirmative case, using Observation \ref{O2}, we can
determine $Y$ in linear-time and so $B=V\setminus (W\cup Y)$.
Therefore, we construct a 3-coloring $(B,Y,W)$ in this way and
then test the validity of $(B,Y,W)$ by checking the conditions of
Observations \ref{O1}, \ref{O2} and \ref{O3}. All these
computations can be done in linear-time.$_\triangle$

\begin{theorem}\label{1black}
Given a connected graph $G$ and $Y\subset V$, there is a
linear-time algorithm to verify if there exists some perfect edge
dominating set whose associated 3-coloring is $(B=\{b\},Y,W)$. Moreover, the algorithm can find a vertex $b$ for which the sum of
the weight of its incident edges is minimum.
\end{theorem}

\proof Clearly, if such perfect edge dominating set exists, it
verifies
\begin{enumerate}
\renewcommand{\labelenumi}{\roman{enumi}.}
\item $G[Y]$ is a graph with maximum degree 1 and at least one
vertex has degree 0. \item $V\setminus Y$ is a independent set.
\item If $\{s_1,s_2,...,s_k\}\subseteq Y$ are the vertices with
degree $0$ in $G[Y]$, and $\{y_1,y_2,...,y_l\}\subset Y$ are the
vertices of degree $1$, the black vertex belongs to
$\bigcap_{i=1}^{k}N(s_i) \setminus \bigcup_{j=1}^{l}N(y_j)$.
\end{enumerate}

Therefore, we check conditions (i) and (ii), and construct
$A=\bigcap_{i=1}^{k}N(s_i) \setminus \bigcup_{j=1}^{l}N(y_j)$ in
linear-time. Note that all vertices in $A$ are equivalent. If
$A\neq\emptyset$, each $b\in A$ generates a valid coloring
$(B,Y,W), B=\{b\}, W=V\setminus (B\cup Y)$. Condition (ii) and
$|B|=1$ imply that $(B,Y,W)$ satisfies the conditions of
Observations \ref{O1} and \ref{O3}. Condition (i) and $b\in A$
imply that $(B,Y,W)$ satisfies the condition of Observation
\ref{O2}. So, the validity of $(B,Y,W)$  holds.$_\triangle$\\

The next theorem refers to finding an efficient edge dominating
set of a graph, given a fixed size vertex dominating set of it.

\begin{theorem}\label{TDIMDomination}
\cite{Li-Mi-Sz-13a} Given a graph $G$ and a vertex dominating set
of fixed size of $G$, there is a linear-time algorithm to solve
the minimum weight efficient edge domination problem for $G$.
\end{theorem}

The following results relate trivial perfect edge dominating
sets of a graph and the existence of vertex dominating complete
subgraphs of certain sizes.

\begin{observation}\label{TBlackDominatingSet}
Given a graph $G$, a dominating set $D$ of $G$, a perfect edge dominating set $P$ of $G$ and $(B,Y,W)$ the 3-coloring associated to $P$, if $D\subseteq B$ then $P$ is the trivial perfect edge dominating set.
\end{observation}

This observation is a direct consequence of Observation \ref{O3}.

\begin{corollary}\label{C1}
Given a connected graph $G$, if there is a dominating $K_p$ with
$p\geq 4$ then $G$ has exactly one perfect edge dominating set $P$
and $P$ is trivial.
\end{corollary}

\proof Suppose that there is a non-trivial perfect edge dominating
set $P$. By Observation \ref{O4}, all vertices of $K_p$ must be
black and by Observation \ref{TBlackDominatingSet}, $P$ must be
trivial and this is a contradiction. $_\triangle$

\begin{corollary}\label{C2} Given a connected $P_5$-free graph
$G$, if $G$ admits some non-trivial perfect edge dominating set
$P$ then $G$ has dominating induced $P_3$ or a dominating $K_3$.
\end{corollary}

\proof If $G$ has at most 2 vertices, then $G$ has only one
perfect edge dominating set which is trivial. Therefore, $G$ has
at least 3 vertices. Now, suppose that $G$ has neither dominating
induced $P_3$ nor dominating $K_3$. By Theorem \ref{TTuza}, $G$
must have some dominating $K_p$ with $p\ne 3$. In case $p\geq 4$,
by Corollary \ref{C1},  $G$ has no non-trivial perfect edge
dominating set which is a contradiction. Therefore, $p\leq 2$ . As
$G$ is connected with at least 3 vertices, it is always possible
to add more vertices to $K_p$ to form a dominating $K_3$ or a
dominating induced $P_3$. Again, a contradiction.
$_\triangle$

\begin{corollary}\label{C3}
Given a graph $G$, if $G$ has some vertex dominating $K_1$ or
$K_3$ then $G$ has no proper edge dominating sets.
\end{corollary}

\proof 
First, assume $G$ has a dominating $K_1=\{u\}$ which means that $u$ 
is a universal vertex. Suppose that $P$ is a proper edge dominating 
set. Let $v$ be any black vertex. By Observation \ref{TBlackDominatingSet}, 
$v$ cannot be a universal vertex. As $u$ is a neighbor of
$v$, then $u$ is a yellow vertex because it is a universal vertex. 
As $v$ is a black vertex, $v$ has another non-white neighbor $w$. 
Moreover, $w$ is adjacent to the universal vertex $u$. This 
contradicts that $u$ is a yellow vertex.

Next, suppose $G$ contains a dominating $K_3$. Suppose that $G$
has some proper edge dominating set $P$ and $(B,Y,W)$, its
associated 3-coloring, with $B,Y,W\ne\emptyset$. By Observation
\ref{O5}, (i) all vertices of the dominating $K_3$ are black or
(ii) exactly two of them are yellow vertices and the other vertex
is white. In case (i), by Observation \ref{TBlackDominatingSet}, $P$
must be trivial which is a contradiction. In case (ii), there are
no black vertices as consequence of Observations \ref{O2} and
\ref{O3}. Again, a contradiction.$_\triangle$\\

Finally, the last theorem relates the existence of proper edge
dominating sets to special colorings of a dominating $P_3$.

\begin{figure}
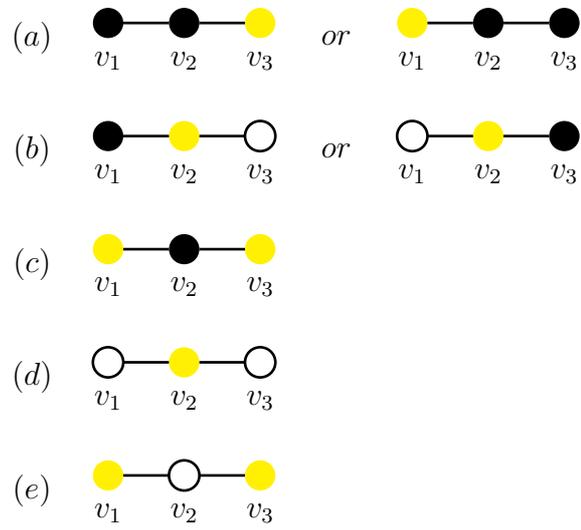

\center
\vspace{0.5cm}
\begin{pgfpicture}{0cm}{0cm}{8cm}{6.9cm}
	\pgfsetlinewidth{1pt}

	\pgfnodecircle{v1}[fill]{\pgfxy(1,1+6)}{2mm}
	\pgfnodecircle{v2}[fill]{\pgfxy(2,1+6)}{2mm}
	\color{yellow}
	\pgfnodecircle{v3}[fill]{\pgfxy(3,1+6)}{2mm}
	\color{black}
	\pgfnodeconnline{v1}{v2}
	\pgfnodeconnline{v2}{v3}			
	\pgfputat{\pgfxy(0,0.8+6)}{\pgfbox[center,center]{\large $(a)$}}
	\pgfputat{\pgfxy(1,1-0.5+6)}{\pgfbox[center,center]{\large $v_1$}}
	\pgfputat{\pgfxy(2,1-0.5+6)}{\pgfbox[center,center]{\large $v_2$}}
	\pgfputat{\pgfxy(3,1-0.5+6)}{\pgfbox[center,center]{\large $v_3$}}

	\pgfnodecircle{v3}[fill]{\pgfxy(3+4,1+6)}{2mm}
	\pgfnodecircle{v2}[fill]{\pgfxy(2+4,1+6)}{2mm}
	\color{yellow}
	\pgfnodecircle{v1}[fill]{\pgfxy(1+4,1+6)}{2mm}
	\color{black}
	\pgfnodeconnline{v1}{v2}
	\pgfnodeconnline{v2}{v3}			
	\pgfputat{\pgfxy(0+4,0.8+6)}{\pgfbox[center,center]{\large $or$}}
	\pgfputat{\pgfxy(1+4,1-0.5+6)}{\pgfbox[center,center]{\large $v_1$}}
	\pgfputat{\pgfxy(2+4,1-0.5+6)}{\pgfbox[center,center]{\large $v_2$}}
	\pgfputat{\pgfxy(3+4,1-0.5+6)}{\pgfbox[center,center]{\large $v_3$}}

	\pgfnodecircle{v1}[fill]{\pgfxy(1,1+4.5)}{2mm}
	\color{yellow}
	\pgfnodecircle{v2}[fill]{\pgfxy(2,1+4.5)}{2mm}
	\color{black}
	\pgfnodecircle{v3}[stroke]{\pgfxy(3,1+4.5)}{2mm}
	\pgfnodeconnline{v1}{v2}
	\pgfnodeconnline{v2}{v3}			
	\pgfputat{\pgfxy(0,0.8+4.5)}{\pgfbox[center,center]{\large $(b)$}}
	\pgfputat{\pgfxy(1,1-0.5+4.5)}{\pgfbox[center,center]{\large $v_1$}}
	\pgfputat{\pgfxy(2,1-0.5+4.5)}{\pgfbox[center,center]{\large $v_2$}}
	\pgfputat{\pgfxy(3,1-0.5+4.5)}{\pgfbox[center,center]{\large $v_3$}}
	\pgfnodecircle{v1}[stroke]{\pgfxy(1+4,1+4.5)}{2mm}
	\color{yellow}
	\pgfnodecircle{v2}[fill]{\pgfxy(2+4,1+4.5)}{2mm}
	\color{black}
	\pgfnodecircle{v3}[fill]{\pgfxy(3+4,1+4.5)}{2mm}
	\pgfnodeconnline{v1}{v2}
	\pgfnodeconnline{v2}{v3}			
	\pgfputat{\pgfxy(0+4,0.8+4.5)}{\pgfbox[center,center]{\large $or$}}
	\pgfputat{\pgfxy(1+4,1-0.5+4.5)}{\pgfbox[center,center]{\large $v_1$}}
	\pgfputat{\pgfxy(2+4,1-0.5+4.5)}{\pgfbox[center,center]{\large $v_2$}}
	\pgfputat{\pgfxy(3+4,1-0.5+4.5)}{\pgfbox[center,center]{\large $v_3$}}

	\color{yellow}
	\pgfnodecircle{v1}[fill]{\pgfxy(1,1+3)}{2mm}
	\pgfnodecircle{v3}[fill]{\pgfxy(3,1+3)}{2mm}
	\color{black}
	\pgfnodecircle{v2}[fill]{\pgfxy(2,1+3)}{2mm}
	\pgfnodeconnline{v1}{v2}
	\pgfnodeconnline{v2}{v3}			
	\pgfputat{\pgfxy(0,0.8+3)}{\pgfbox[center,center]{\large $(c)$}}
	\pgfputat{\pgfxy(1,1-0.5+3)}{\pgfbox[center,center]{\large $v_1$}}
	\pgfputat{\pgfxy(2,1-0.5+3)}{\pgfbox[center,center]{\large $v_2$}}
	\pgfputat{\pgfxy(3,1-0.5+3)}{\pgfbox[center,center]{\large $v_3$}}

	\pgfnodecircle{v1}[stroke]{\pgfxy(1,1+1.5)}{2mm}
	\color{yellow}
	\pgfnodecircle{v2}[fill]{\pgfxy(2,1+1.5)}{2mm}
	\color{black}
	\pgfnodecircle{v3}[stroke]{\pgfxy(3,1+1.5)}{2mm}
	\pgfnodeconnline{v1}{v2}
	\pgfnodeconnline{v2}{v3}			
	\pgfputat{\pgfxy(0,0.8+1.5)}{\pgfbox[center,center]{\large $(d)$}}
	\pgfputat{\pgfxy(1,1-0.5+1.5)}{\pgfbox[center,center]{\large $v_1$}}
	\pgfputat{\pgfxy(2,1-0.5+1.5)}{\pgfbox[center,center]{\large $v_2$}}
	\pgfputat{\pgfxy(3,1-0.5+1.5)}{\pgfbox[center,center]{\large $v_3$}}

	\color{yellow}
	\pgfnodecircle{v1}[fill]{\pgfxy(1,1)}{2mm}
	\pgfnodecircle{v3}[fill]{\pgfxy(3,1)}{2mm}
	\color{black}
	\pgfnodecircle{v2}[stroke]{\pgfxy(2,1)}{2mm}
	\pgfnodeconnline{v1}{v2}
	\pgfnodeconnline{v2}{v3}			
	\pgfputat{\pgfxy(0,0.8)}{\pgfbox[center,center]{\large $(e)$}}
	\pgfputat{\pgfxy(1,1-0.5)}{\pgfbox[center,center]{\large $v_1$}}
	\pgfputat{\pgfxy(2,1-0.5)}{\pgfbox[center,center]{\large $v_2$}}
	\pgfputat{\pgfxy(3,1-0.5)}{\pgfbox[center,center]{\large $v_3$}}
\end{pgfpicture}
\caption{Possible valid colorings of dominating induced $P_3$}\label{P3}
\end{figure}

\begin{theorem}\label{ppal}
Given a connected $P_5$-free graph $G$, if $G$ admits some proper
perfect edge dominating set $P$, with associated 3-coloring
$(B,Y,W)$, then $G$ has a vertex dominating $P_3$ (formed by
vertices $v_1,v_2$ and $v_3$) with one of the possible
combinations of colors of Figure \ref{P3}.
\end{theorem}

\begin{figure}
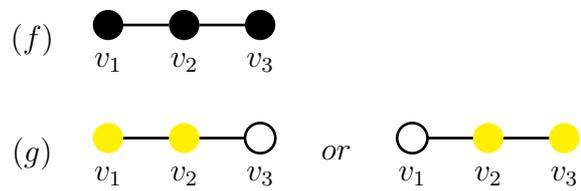

\center
\begin{pgfpicture}{0cm}{0cm}{8cm}{2.85cm}
    \pgfsetlinewidth{1pt}

   \pgfnodecircle{v1}[fill]{\pgfxy(1,1+1.5)}{2mm}
   \pgfnodecircle{v2}[fill]{\pgfxy(2,1+1.5)}{2mm}
   \pgfnodecircle{v3}[fill]{\pgfxy(3,1+1.5)}{2mm}
   \pgfnodeconnline{v1}{v2}
   \pgfnodeconnline{v2}{v3}
   \pgfputat{\pgfxy(0,0.8+1.5)}{\pgfbox[center,center]{\large $(f)$}}
   \pgfputat{\pgfxy(1,1-0.5+1.5)}{\pgfbox[center,center]{\large $v_1$}}
   \pgfputat{\pgfxy(2,1-0.5+1.5)}{\pgfbox[center,center]{\large $v_2$}}
   \pgfputat{\pgfxy(3,1-0.5+1.5)}{\pgfbox[center,center]{\large $v_3$}}

    \color{yellow}
    \pgfnodecircle{v1}[fill]{\pgfxy(1,1)}{2mm}
    \pgfnodecircle{v2}[fill]{\pgfxy(2,1)}{2mm}
    \color{black}
    \pgfnodecircle{v3}[stroke]{\pgfxy(3,1)}{2mm}
    \pgfnodeconnline{v1}{v2}
    \pgfnodeconnline{v2}{v3}
    \pgfputat{\pgfxy(0,0.8)}{\pgfbox[center,center]{\large $(g)$}}
    \pgfputat{\pgfxy(1,1-0.5)}{\pgfbox[center,center]{\large $v_1$}}
    \pgfputat{\pgfxy(2,1-0.5)}{\pgfbox[center,center]{\large $v_2$}}
    \pgfputat{\pgfxy(3,1-0.5)}{\pgfbox[center,center]{\large $v_3$}}

    \color{yellow}
    \pgfnodecircle{v2}[fill]{\pgfxy(2+4,1)}{2mm}
    \pgfnodecircle{v3}[fill]{\pgfxy(3+4,1)}{2mm}
    \color{black}
    \pgfnodecircle{v1}[stroke]{\pgfxy(1+4,1)}{2mm}
    \pgfnodeconnline{v1}{v2}
    \pgfnodeconnline{v2}{v3}
    \pgfputat{\pgfxy(0+4,0.8)}{\pgfbox[center,center]{\large $or$}}
    \pgfputat{\pgfxy(1+4,1-0.5)}{\pgfbox[center,center]{\large $v_1$}}
    \pgfputat{\pgfxy(2+4,1-0.5)}{\pgfbox[center,center]{\large $v_2$}}
    \pgfputat{\pgfxy(3+4,1-0.5)}{\pgfbox[center,center]{\large $v_3$}}
\end{pgfpicture}
\caption{Invalid colorings of dominating induced $P_3$}
\label{P32}
\end{figure}

\proof Clearly, by Corollaries \ref{C2} and \ref{C3} $G$ has a
dominating induced $P_3$. By Observations \ref{O1}, \ref{O2} and
\ref{O3}, there exists neither two adjacent white vertices, nor yellow
vertices with at least two non-white neighbors nor white vertices
with black neighbors. As a consequence, the possible colors of the
dominating induced $P_3$ must match to some combinations of Figure
\ref{P3} or some of Figure \ref{P32}. If the combination (f) of
Figure \ref{P32} is matched  then all vertices of the
dominating induced $P_3$ are black vertices and by Observation
\ref{TBlackDominatingSet}, $P$ is a trivial perfect edge dominating
set leading to a contradiction. If the combination (g) of Figure
\ref{P32} is matched, two adjacent vertices of the dominating
induced $P_3$ are yellow vertices and the other one is a white
vertex. None of these three vertices can have black neighbors
which implies $B=\emptyset$ and $P$ to be an efficient dominating
set. Again, this is a contradiction. Therefore, the only valid
options are those of Figure \ref{P3}.$_\triangle$

\subsection{The algorithm}

Let $G$, $|V(G)| > 1$, be an arbitrary connected graph. The
proposed algorithm, along the process,  constructs a set
${\mathcal E}$ containing a few candidates for the minimum
weight perfect edge dominating set of $G$. At the end, it
selects the least of them and returns the minimum edge dominating
set of $G$,

\begin{enumerate}
\item Define ${\mathcal E} := \{E(G)\}$

\item Find a principal vertex of $G$. If no such vertex exists
then return an induced $P_5$ and stop.

\item Using the principal vertex $v$, find (i) an induced $P_5$,
or (ii) a dominating $K_p$, or (iii) a dominating $P_3$.

\item Case (i): An induced $P_5$ is found. Then return it and
stop.

\item Case (ii): A dominating $K_p$ is found. If $p \geq 4$ then
return $E(G)$ and stop. Otherwise, $p \leq 3$ and using such
dominating $K_p$, find a minimum weight efficient edge
dominating set and, if it exists, include it in ${\mathcal E}$. If
$K_p = \{v_1,v_2\}$ and $N(v_1)\cap N(v_2)=\emptyset$ then transform 
the dominating $K_2$ into a dominating
$P_3$ by adding a third vertex to it.

\item Case (iii): A dominating $P_3$ is found. First, again using
such a vertex dominating set, find a minimum weight efficient edge dominating
set of $G$, and include it in ${\mathcal E}$, if it exists. Then
look for a proper perfect edge dominating set of $G$, by
considering every possible coloring of the $P_3$, according to
Figure \ref{P3}. In each of the cases (a)-(b) ((c)-(e)), below, the algorithm
determines at most two (one) proper perfect edge dominating sets (set), which
are (is) then included in ${\mathcal E}$.

\begin{description}
\item[(a)] Without loss of generality, $v_3$ is the yellow vertex
of the dominating induced $P_3$. Clearly, $W=N(v_3)\setminus
\{v_2\}$ and we can apply the linear-time algorithm of Theorem
\ref{algo} to determine the proper perfect edge dominating set
whose associated 3-coloring is $(B,Y,W)$. Include it in ${\mathcal
E}$, if it exists. \item[(b)] Without loss of generality, $v_1$ is
the black vertex of the dominating induced $P_3$. We can applied
the same technique of (a) using $W=N(v_2)\setminus \{v_1\}$.

\item[(c)] Again, apply the same technique of (a) using
$W=(N(v_1)\cup N(v_3))\setminus \{v_2\}$.

\item[(d)] In this case, $v_1$ and $v_3$ have only yellow
neighbors. As we are looking for proper perfect edge dominating
sets, there is some black vertex somewhere. Clearly, $v_2$ must
have exactly one black neighbor and the other neighbors of $v_2$
are white vertices. So, $Y=N(v_1)\cup N(v_3)$, $|B|=1$, and we can
apply the linear-time algorithm of Theorem \ref{1black} to
determine the proper perfect edge dominating set whose associated
3-coloring is $(B,Y,W)$ with least weight. Again, if successful include it in
${\mathcal E}$.

\item[(e)] In this case, $v_1$ ($v_3$) can have at most one black
neighbor. As we are looking for proper perfect edge dominating
sets, there is at least one black vertex. First, we assume that
there are some triangles using edges of the dominating induced
$P_3$ (the number of these triangles is exactly $|N(v_1)\cap
N(v_2)|+|N(v_2)\cap N(v_3)|$ which can be computed in
linear-time). If there are at least 2 triangles using the same
edge, without loss of generality $\{v_1,v_2,x\}$ and
$\{v_1,v_2,x'\}$, then the coloring is invalid. If $\{v_1,v_2,x\}$
and $\{v_2,v_3,x'\}$ are triangles then the coloring is invalid or
it cannot admit black vertices, which means there are no proper
perfect edge dominating sets. The only possibility for the graph
to admit a proper perfect edge dominating set (which implies the
existence of black vertices) is the existence of exactly one triangle
using an edge of the dominating induced $P_3$ and $|B|=1$.
Clearly, the yellow vertices are $Y=N(v_2)$ and we then proceed as
in Case (d). Next, suppose there is no triangle using an edge of
the dominating induced $P_3$. Clearly, every vertex in
$A_{1,3}=N(v_1)\setminus N(v_3)$ has to be adjacent to every
vertex in $A_{3,1}=N(v_3)\setminus N(v_1)$ or there is some
induced $P_5$, which the algorithm returns  and then stops. It is
not hard to see this can be accomplished in linear-time. Suppose
that there is only a black vertex $b$. It has to be adjacent to
$v_1$ and $v_3$, otherwise (without loss of generality,
$bv_1\notin E(G) $) there is an induced $P_5$ formed by
$y,v_1,v_2,v_3,b$ where $y$ is a yellow neighbor of $v_1$. Then
the yellow vertices are $Y=N(v_2)$ and we proceed as in
(d). The last case is that there are two black vertices $b_1,b_3$
where $b_1\in A_{1,3}$ and $b_3\in A_{3,1}$, and the coloring is
invalid if $|A_{1,3}|>1$ or $|A_{3,1}|>1$. The algorithm will
check if $(B=\{b_1,b_3\},Y=N(v_2),W=V\setminus(B\cup Y))$ is a
valid coloring, and include in ${\mathcal E}$ the corresponding
edge dominating set. All these computations can be done in
linear-time.
\end{description}

\item Select the least weight perfect edge dominating set of
${\mathcal E}$, return it and stop.
\end{enumerate}

The correctness and linear time complexity of the algorithm
follows directly from the propositions formulated in the previous
subsections.

To summarise we state the main result of this section.

\begin{theorem}
The weighted perfect edge domination problem can be solved for 
$P_5$-free graphs in linear time in a robust way.
\end{theorem}



%
%

\end{document}